\documentclass[a4paper,twocolumn,10pt]{quantumarticle}
\pdfoutput=1
\usepackage[numbers,sort&compress]{natbib}
\usepackage{adjustbox}
\usepackage{amssymb,amsmath,amstext,amsthm}
\DeclareFontFamily{U}{mathx}{}
\DeclareFontShape{U}{mathx}{m}{n}{<-> mathx10}{}
\DeclareSymbolFont{mathx}{U}{mathx}{m}{n}
\DeclareMathAccent{\widecheck}{0}{mathx}{"71}
\usepackage{dsfont}
\usepackage{graphicx}
\usepackage{bm}
\usepackage{algorithm,algpseudocode}
\usepackage{appendix}
\usepackage[T1]{fontenc}
\usepackage{physics}
\usepackage{xcolor}
\usepackage{hyperref}
\hypersetup{colorlinks=true,citecolor=blue,linkcolor=magenta}
\usepackage{tikz-network}
\usepackage{orcidlink}
\makeatletter
\def\rt@tablename{table}
\g@addto@macro\@floatboxreset{\ifx\@captype\rt@tablename\small\fi}
\newenvironment{ruledtabular}{%
  \setlength{\tabcolsep}{3pt}%
  \expandafter\let\expandafter\rt@tabstar\csname tabular*\endcsname
  \let\rt@endtab\endtabular
  \def\tabular##1{\rt@tabstar{\linewidth}{@{\extracolsep{\fill}}##1}\hline\hline}%
  \def\endtabular{\hline\hline\rt@endtab}}{}
\renewcommand{\widetext}{\onecolumn}
\makeatother

\newcommand{\SU}[1]{\ensuremath{\mathrm{SU}(#1)}}
\newcommand{\U}[1]{\ensuremath{\mathrm{U}(#1)}}
\newcommand{\cg}[6]{\mathcal{C}^{#5 #6}_{#1 #2;\, #3 #4}}

\newtheorem{theorem}{Theorem}
\newtheorem{definition}{Definition}

\newtheorem{proposition}{Proposition}
\newtheorem{lemma}{Lemma}
\newtheorem{corollary}{Corollary}
\newtheorem*{informaltheorem}{Theorem~\ref{thm:preparation} (informal)}

\newcommand{\quantummotionaus}{Quantum Motion, 4 Cornwallis St, Eveleigh NSW 2015, Australia}
\newcommand{\quantummotionuk}{Quantum Motion, 9 Sterling Way, London N7 9HJ, United Kingdom}
\newcommand{\unimelb}{School of Physics, The University of Melbourne, Parkville, VIC 3010, Australia}
\newcommand{\oxfordmath}{Mathematical Institute, University of Oxford, Woodstock Road, Oxford OX2 6GG, United Kingdom}

\begin{document}

\title{Spin-adapted state preparation for quantum chemistry}

\author{Matthew~L.~Sims-Goh\,\orcidlink{0000-0002-7478-4026}}
\email{matt.sims-goh@quantummotion.tech}
\affiliation{\quantummotionaus}
\affiliation{\unimelb}

\author{Fredrik Hasselgren \,\orcidlink{0009-0007-9273-4480}}
\affiliation{\quantummotionuk}
\affiliation{\oxfordmath}

\author{Hamza Jnane\,\orcidlink{0000-0002-0713-3354}}
\affiliation{\quantummotionuk}

\begin{abstract}
Quantum phase estimation for strongly-correlated molecules requires an initial state of high overlap with the true ground state, and preparing one is a significant fault-tolerant cost. Matrix product states (MPS) produced by the density matrix renormalisation group (DMRG) are a leading choice, yet existing circuits treat their tensors as dense or do not use the full symmetries present in molecular electronic Hamiltonians. In quantum chemistry, DMRG also exploits total spin, a non-abelian symmetry, storing one amplitude per spin multiplet instead of one per state. We show that the reduced tensors of such a spin-adapted MPS, labelled by the steps of the unitary group approach, form an abelian-symmetric MPS, which existing block-sparse preparation protocols can load at a cost set by the multiplet count. One state-independent rotation per site, determined by a Clebsch-Gordan coefficient, then returns the state to the occupation basis. We provide resource estimates for our circuits on singlet ground states of active space models of iron-sulphur clusters, finding improvements in Toffoli count up to $135\times$ compared to dense methods and $4.9\times$ compared to block-sparse methods applied to the same states, with up to $3.7\times$ fewer ancilla qubits required.
\end{abstract}
\maketitle

\section{Introduction\label{sec:introduction}}
As utility-scale quantum computers draw closer to reality, the search for quantum applications of genuine economic and scientific value has assumed new urgency. Of these, studying ground states of chemical and material systems appears the most promising --- with well-known quantum phase estimation (QPE) procedures that can in principle achieve an exponential speedup~\cite{kitaev1995quantum,abrams1999quantum,aspuru2005simulated}, it is widely hoped that \emph{in silico} study with quantum processors can enhance drug discovery~\cite{blunt2022perspective,goings2022reliably,santagati2024drug}, unlock efficient nitrogen-fixation pathways~\cite{reiher2017elucidating,berry2019qubitization,lee2021even,von2021quantum}, and accelerate the design of next-generation batteries~\cite{rice2021quantum,kim2022fault,delgado2022simulating}.
However, the success of such a calculation hinges on whether one can supply it with a sufficiently good initial state~\cite{ollitrault2024enhancing,lee2023evaluating,tubman2018postponing,marti2025spin,fomichev2024initial,berry2025rapid} --- indeed, the success probability of QPE scales as the squared overlap of the initial state with the true ground state~\cite{abrams1999quantum}. Despite being a critical bottleneck for QPE, resource analyses often assume that such a state is available at perfect overlap for a cost far below that of the phase estimation itself~\cite{lee2021even,fomichev2024initial,berry2025rapid,rupprecht2026faster}. Whether a state of sufficient overlap can be found at all may determine whether an exponential advantage is achievable in quantum chemistry~\cite{lee2023evaluating}. After recent reductions in the cost of phase estimation, preparing such a state can cost as much as the phase estimation itself~\cite{low2025fast,rupprecht2026faster}.

Generic quantum states are exponentially costly to prepare~\cite{knill1995approximation,mottonen2004transformation,shende2005synthesis}, so efficient quantum state preparation must rely on some hidden underlying structure, typically related to some notion of classical simulability~\cite{tubman2018postponing,marti2025spin,fomichev2024initial,berry2025rapid,rupprecht2026faster,sims2026multivariate,holmes2020efficient,gonzalez2024efficient,iaconis2024quantum,bohun2026entanglement,creevey2025scalable,hasselgren2026breaking,ran2020encoding,huggins2025efficient,schoen2005sequential,grover2002creating,rattew2022preparing,mcardle2022quantum,moosa2023linear,marin2023quantum,mori2024efficient,rosenkranz2025quantum,kivlichan2018quantum,goh2025lie,barligea2026enabling,rupprecht2026sparse,burkat2025quantum}. A delicate balance must be maintained: the simplest of these have exponentially-vanishing success probabilities in QPE rendering them unsuitable~\cite{lee2023evaluating,fomichev2024initial}, but loading excessively complicated initial states can exceed the cost of the quantum computation itself~\cite{aaronson2015read,dalzell2025quantum}. Tensor networks are an extremely useful choice for quantum state preparation~\cite{schoen2005sequential,holmes2020efficient,ran2020encoding,gonzalez2024efficient,iaconis2024quantum,fomichev2024initial,huggins2025efficient,berry2025rapid,bohun2026entanglement,rupprecht2026faster,sims2026multivariate,hasselgren2026breaking,creevey2025scalable}, in part because their bond dimension $\chi$ provides a tunable `dial' which one can achieve this balance by tweaking. Matrix product states (MPS) are the simplest and most popular tensor network ansatz, and have been highly successful in the classical simulation of quantum~\cite{vidal2003efficient,vidal2004efficient,haegeman2011time,haegeman2016unifying,paeckel2019time,white1992density,white1993density,schollwock2005density,schollwock2011density,mcculloch2007density,weichselbaum2012non,sharma2012spin,keller2016spin,singh2012tensor} and classical~\cite{gourianov2022quantum,ye2022quantum,lee2026quantum} systems; they are accordingly a leading vector for quantum state preparation of both ground states~\cite{ran2020encoding,fomichev2024initial,huggins2025efficient,berry2025rapid,rupprecht2026faster,hasselgren2026breaking} and classical data~\cite{holmes2020efficient,gonzalez2024efficient,iaconis2024quantum,bohun2026entanglement,creevey2025scalable,hasselgren2026breaking}.

\begin{figure*}[t]
\centering
\includegraphics[width=\textwidth]{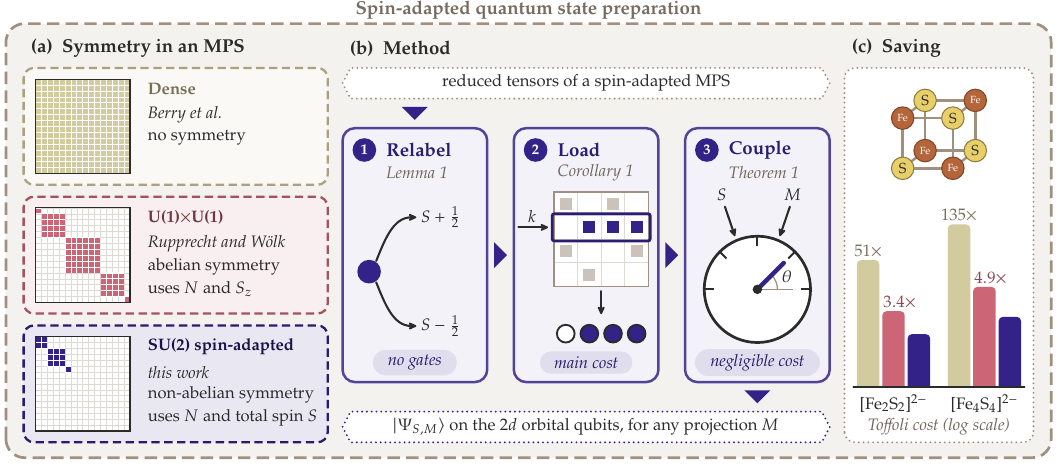}
\caption{\label{fig:overview}\textbf{Overview of this work.} \textbf{(a)} One site matrix of the same matrix product state in three descriptions, with the stored entries filled. Without symmetry the matrix is dense~\cite{fomichev2024initial,berry2025rapid} (gold); the abelian symmetries of particle number $N$ and spin projection $S_z$ make it block-sparse~\cite{rupprecht2026faster} (red); total spin $S$, a non-abelian symmetry, leaves one entry for each multiplet in place of one for each of its $2S+1$ states (blue, ours). \textbf{(b)} The three stages of our construction: a classical relabelling of the reduced tensors (Lemma~\ref{lemma:mapping}), enabling one to use the block-sparse loading of Ref.~\cite{rupprecht2026faster} (Corollary~\ref{corollary:loading}), and one rotation at each site, by an angle set by the spin $S$ and its projection $M$ carried along the bonds (Theorem~\ref{thm:preparation}). \textbf{(c)} The savings in Toffoli cost of preparing the singlet ground states of two iron-sulphur clusters from the three MPS descriptions, with the factors by which the dense and the $\U1\times\U1$ preparation costs exceed ours (Table~\ref{tab:costs}).}
\end{figure*}

Ground states of quantum systems can be approximated with the density matrix renormalisation group (DMRG)~\cite{white1992density,white1993density,schollwock2005density,mcculloch2007density,schollwock2011density,sharma2012spin,weichselbaum2012non,keller2016spin}, which has been highly successful in quantum chemistry~\cite{chan2011density,sharma2012spin,sharma2014low,keller2016spin,baiardi2020density,zhai2023block2,menczer2024tensor,menczer2024parallel,xiang2024distributed,legeza2025orbital}.
Much of this success rests on symmetry. Molecular electronic Hamiltonians conserve the electron number, the spin projection and the total spin, and often have point-group symmetry as well; fully exploiting these symmetries reduces the computational cost enormously~\cite{mcculloch2007density,weichselbaum2012non,singh2012tensor,sharma2012spin,keller2016spin}. DMRG libraries that exploit the total spin are called `spin-adapted', and the most advanced quantum chemistry calculations would not be possible without them.
MPS-based state preparation, meanwhile, lags behind greatly in these optimisations: the vast majority of routines load MPS in their full, symmetry-unconstrained form, even in cases where symmetries compress the meaningful information by orders of magnitude~\cite{schoen2005sequential,ran2020encoding,fomichev2024initial,berry2025rapid,hasselgren2026breaking}.
While symmetry can be projectively imposed after loading~\cite{khinevich2026symmetry}, this does not reduce the cost of loading the state itself.
The cost of preparing a state should ultimately reflect only the number of parameters that the classical calculation stores for it, but the first exploitation of symmetry for MPS state preparation that we are aware of was several months ago by Rupprecht and W\"{o}lk~\cite{rupprecht2026faster}, who used abelian-symmetry-induced block-sparsity to reduce the loading costs accordingly. This enables one to avoid incurring costs associated with information that is redundant under abelian symmetries such as particle number and spin projection, but is incompatible with non-abelian symmetries due to their more sophisticated representation theory and fusion rules, necessitating an inflation of the underlying state's representation to map it to a quantum circuit. In this work, we complete this progression towards exploiting the full symmetries of quantum chemistry Hamiltonians by exploiting their $\SU2$-induced non-abelian symmetry, allowing spin-adapted MPS to be prepared at a cost reflective of their actual classical size. Our main result is stated informally as follows:
\begin{informaltheorem}
Let a spin-adapted MPS on $d$ orbitals represent the states $\ket{\Psi_{SM}}$ of total spin $S$ and projection $M=-S,\dots,S$. A single quantum circuit can prepare $\ket{\Psi_{SM}}$ for any $M$, selected by the initial value of a register of $\mathcal{O}(\log d)$ qubits. Its Toffoli cost inherits that of the block-sparse preparation of Ref.~\cite{rupprecht2026faster} but with blocks counting multiplets instead of states (i.e. $2S'+1$ times fewer amplitudes per multiplet of spin $S'$), plus a relatively small overhead at each site that does not depend on the entries of the tensors.
\end{informaltheorem}
We give an overview of our work in Figure~\ref{fig:overview}; Figure~\ref{fig:overview}(a) illustrates how the information required to represent an MPS for a quantum chemistry problem can vary dramatically depending on the level of symmetry exploited.
The obstacle to extending block-sparse preparation to spin-adapted MPS is that total spin, unlike particle number, is not additive: a singly occupied orbital can either raise or lower the spin of the orbitals it is coupled to, so the state of a site no longer determines how the symmetry label on its bond changes.
Our construction (Figure~\ref{fig:overview}(b)) removes the obstacle by labelling each site with which of the two happens.
This label is the \emph{step} of the unitary group approach~\cite{paldus1974group,shavitt1977graph,shavitt1978matrix}, in which quantum chemists have long written their spin-adapted basis states. In terms of steps, the tensors that a spin-adapted MPS stores, its \emph{reduced tensors}, form an abelian-symmetric MPS, which the block-sparse preparation of Ref.~\cite{rupprecht2026faster} loads without modification. A state-independent rotation at each site then returns the state to the occupation basis; together these rotations form the inverse of the quantum Paldus transform of Ref.~\cite{burkat2025quantum}.

We demonstrate the practical savings of our circuit construction on DMRG approximations of the singlet ground states (total spin $S=0$) of the iron-sulphur clusters [Fe$_2$S$_2$]$^{2-}$ and [Fe$_4$S$_4$]$^{2-}$, finding savings of up to $4.9\times$ fewer Toffoli gates compared to Ref.~\cite{rupprecht2026faster} and up to $135\times$ fewer Toffoli gates than the protocol of Ref.~\cite{berry2025rapid}. The saving persists even when the states are truncated to equal fidelity, and it grows with the share of the Schmidt weight that multiplets of large spin carry across the bonds.

The paper is organised as follows. Section~\ref{sec:preliminaries} presents a pedagogical review of the necessary background, building up MPS with abelian and $\SU2$ symmetries, the spin-adapted MPS and the unitary group approach, and fixes the conventions for fermions and for the coupling of spins on which the circuit depends. Section~\ref{sec:results} contains the circuit construction, its cost, and our main theoretical results and proofs. Section~\ref{sec:numerics} presents and analyses resource estimates for concrete problem instances in quantum chemistry, and Section~\ref{sec:conclusion} concludes.

\section{Preliminaries}\label{sec:preliminaries}

\subsection{Setup}\label{sec:problem-setup}
Throughout, we consider systems of $d$ spatial orbitals, ordered as $i=1,\dots,d$, each carrying spin-orbitals $(i,\uparrow)$ and $(i,\downarrow)$ such that the Fock space consists of $2d$ fermionic modes. We consider mapping to the register of a quantum computer via a Jordan-Wigner mapping in the interleaved order $(1,\uparrow), (1,\downarrow),(2,\uparrow),(2,\downarrow)$. We will be concerned with Hamiltonians possessing $\U1\times\SU2$ symmetry (i.e. commuting with the number operator $\hat{N}$ and total spin operator $\hat{S}^2$), most commonly the spin-free electronic Hamiltonians of quantum chemistry
\begin{equation}
    H=\sum_{ij,\mu} h_{ij}a^\dagger_{i\mu} a_{j\mu} + \frac{1}{2}\sum_{ijkl,\mu\nu} v_{ij,kl} a^\dagger_{i\mu} a^\dagger_{j\nu} a_{l\nu} a_{k\mu},\label{eqn:chemistry-hamiltonian}
\end{equation}
where $\mu,\nu\in\{\uparrow,\downarrow\}$ label the spin of the fermionic spin-orbitals, $h_{ij}$ are the one-electron integrals and $v_{ij,kl}$ the two-electron integrals (carrying no symmetry-breaking $\mu$ or $\nu$ dependence), and physicists' index convention $v_{ij,kl}=\bra{ij}r_{12}^{-1}\ket{kl}=(ik \vert jl)$. Since these Hamiltonians commute with $\hat{N}$, $\hat{S}^2$, and $\hat{S}_z$, their eigenstates can be labelled by quantum numbers $(N, S, M)$, i.e. $N$ electrons, with total spin $S$, and spin projection $M$.
Abelian point-group symmetries can also be exploited in our framework, as we discuss in Section~\ref{sec:loading}.

We consider the task of efficiently loading the ground states of such Hamiltonians into the register of a quantum computer. Following previous works~\cite{fomichev2024initial,berry2025rapid,rupprecht2026faster}, we use the MPS representation, which parametrises a quantum state by a network of $d$ tensors $\{M_i\}$.
\begin{definition}[Matrix product state (MPS)]\label{def:mps}
Let each site $i \in \{1,\dots,d\}$ carry a local basis $\{\ket{\sigma_i}\}$ of dimension $D$.
An MPS is a state of the form
\begin{equation}\label{eq:mps}
    \ket{\Psi} = \sum_{\sigma_1,\dots,\sigma_d} M^{\sigma_1}_1 M^{\sigma_2}_2 \cdots M^{\sigma_d}_d \ket{\sigma_1, \dots, \sigma_d},
\end{equation}
where $M_i^{\sigma_i}$ is the $\chi_{i-1}\times\chi_i$ matrix obtained by fixing the physical index of the site-$i$ tensor $M_i$ to the value $\sigma_i\in\{1,\dots,D\}$, with $\chi_0=\chi_d=1$. The integers $\chi_i$ are the \emph{bond dimensions}, the maximum of which $\chi=\max_i\chi_i$ is typically referred to as the \emph{bond dimension} of the MPS.
\end{definition}
It is often convenient to represent tensor networks in Penrose graphical notation~\cite{penrose1971applications,bridgeman2017hand}, where the amplitudes of Equation~\ref{eq:mps} may be written
\begin{equation}
\braket{\sigma_1,\dots,\sigma_d}{\Psi}=
\begin{array}{c}
\begin{tikzpicture}
\Vertex[x=0,y=0,label=$M_1$,position=above,size=0.25,color=blue!30]{M1}
\Vertex[x=1,y=0,label=$M_2$,position=above,size=0.25,color=blue!30]{M2}
\Vertex[x=2.5,y=0,label=$M_{d-1}$,position=above,size=0.25,color=blue!30]{Mdm1}
\Vertex[x=3.5,y=0,label=$M_d$,position=above,size=0.25,color=blue!30]{Md}
\draw[line width = 1] (M1) -- ++(0,-0.35) node[below] {\scriptsize$\sigma_1$};
\draw[line width = 1] (M2) -- ++(0,-0.35) node[below] {\scriptsize$\sigma_2$};
\draw[line width = 1] (Mdm1) -- ++(0,-0.35) node[below] {\scriptsize$\sigma_{d-1}$};
\draw[line width = 1] (Md) -- ++(0,-0.35) node[below] {\scriptsize$\sigma_d$};
\Edge[lw=1,position={below},label={\makebox[1em][c]{$\alpha_{d-1}$}}](Mdm1)(Md)
\Edge[lw=1,position={below},label={\makebox[1em][c]{$\alpha_1$}}](M1)(M2)
\Edge[lw=1,label=$\cdots$](M2)(Mdm1)
\end{tikzpicture}
\end{array},
\label{eq:mps-diagram}
\end{equation}
where $\alpha_i=1,\dots,\chi_i$ is the bond index connecting the tensors $M_i$ and $M_{i+1}$. The gauge freedom of tensor networks~\cite{schollwock2011density,bridgeman2017hand} means that we are free to place the tensor network in right-canonical form:
\begin{definition}[Right-canonical form]
    An MPS is said to be in right-canonical form if each tensor $M_i$ satisfies the isometric condition
    \begin{equation}
    \begin{array}{c}
    \begin{tikzpicture}
    \Vertex[x=0,y=0.3,label=$M_i$,position=center,size=0.5,color=blue!30]{M1}
    \Vertex[x=0,y=-0.3,label=$M_i^\dagger$,position=center,size=0.5,color=blue!30]{M2}
    \draw[line width = 1] (M1) -- ++(-0.45,0) node[below] {};
    \draw[line width = 1] (M2) -- ++(-0.45,0) node[below] {};
    \Edge[lw=1,bend=90](M1)(M2)
    \Edge[lw=1](M1)(M2)
    \end{tikzpicture}
    \end{array}
    =\begin{array}{c}
    \begin{tikzpicture}
    \draw[line width=1] (0,0.3) to[out=0,in=0,looseness=1.5] (0,-0.3);
    \end{tikzpicture}
    \end{array},
    \label{eq:right-canonical}
    \end{equation}
    i.e. $\sum_{\sigma_i} M_i^{\sigma_i}(M_i^{\sigma_i})^\dagger=\mathds{1}_{\chi_{i-1}}$. This follows the standard convention of the tensor-network literature~\cite{schollwock2011density,bridgeman2017hand}; of the preceding works on state preparation, Ref.~\cite{rupprecht2026faster} uses the same convention, whereas Refs.~\cite{fomichev2024initial,berry2025rapid} call this form left-canonical. For the avoidance of doubt, this convention places the orthogonality centre on the left, isometries dilate from left to right, and state preparation proceeds as a left-to-right sweep.
\end{definition}

\subsection{Tensor networks with abelian symmetry}\label{sec:abelian-warmup}

As a first step towards exploiting the full symmetries of chemistry Hamiltonians, we consider tensor networks with abelian symmetries, such as particle number conservation. Symmetries of this type were exploited to reduce state preparation costs in Ref.~\cite{rupprecht2026faster}. Abelian symmetries manifest in MPS as block-sparsity, where large sections of the tensors are guaranteed to be zero due to symmetry constraints. Specifically, the symmetry implies a conservation law, and an additive charge is assigned to each index value of both physical and bond legs to track the flow of this conserved quantity (e.g. an integer for a single conserved quantity such as $\hat{N}$, or a vector of integers for multiple conserved quantities). In chemistry, the conserved quantities are often the number of electrons and the total spin projection. Formally:
\begin{definition}[Abelian-symmetric MPS]\label{def:symmetric-mps}
Let $\hat{Q}\equiv\sum_{i=1}^d \hat{Q}_i$ be a conserved charge that is a sum of single-site terms $\hat Q_i$ (e.g.\ $\hat{N}$ or $\hat{S}_z$). Write its partial sum over the sites to the right of bond $i$ as $\hat Q^{>i} = \sum_{k>i} \hat Q_k$, and choose each local basis $\{\ket{\sigma_i}\}$ to be an eigenbasis of $\hat{Q}_i$ with eigenvalues $q(\sigma_i)$.
An MPS possesses a corresponding \emph{abelian symmetry} if each bond index $\alpha_i$ carries a charge $q(\alpha_i)$, the right boundary bond carries $q(\alpha_d)=0$, and every site tensor $M_i$ obeys the \emph{selection rule}
\begin{equation}\label{eq:charge-cons}
    (M_i^{\sigma_i})_{\alpha_{i-1}\alpha_i} = 0 \quad\text{whenever}\quad q(\alpha_{i-1}) \neq q(\sigma_i) + q(\alpha_i).
\end{equation}
\end{definition}
That is, the charge on bond $i$ counts the charge of everything to its right, and each site adds its own charge as one moves leftwards along the chain, summing up to the total charge $Q$ on the left boundary bond.

This accumulation of charge can be seen from how the state is built from its tensors one orbital at a time, from right to left. We denote by $\ket{\alpha_i}$ the state of orbitals $i+1,\dots,d$ obtained by contracting the tensors to the right of bond $i$ with the index of bond $i$ fixed to $\alpha_i$. The block to the right of bond $d$ is empty, with the Fock vacuum as its state, and orbital $i$ is attached to the block on its right as
\begin{equation}\label{eq:abelian-recursion}
    \ket{\alpha_{i-1}} = \sum_{\sigma_i,\alpha_i} (M_i^{\sigma_i})_{\alpha_{i-1}\alpha_i} \ket{\sigma_i}\otimes\ket{\alpha_i},
\end{equation}
where $\ket{\alpha_0}=\ket{\Psi}$ by construction, since all tensors have been contracted.

We define the vacuum as having zero charge, i.e. $q(\alpha_d)=0$. If each right-block state $\ket{\alpha_i}$ has the definite charge $q(\alpha_i)$, i.e. is an eigenstate of $\hat Q^{>i}$ with that eigenvalue, then every non-zero term in the sum~\eqref{eq:abelian-recursion} has charge $q(\sigma_i)+q(\alpha_i)=q(\alpha_{i-1})$, since the selection rule~\eqref{eq:charge-cons} sets all other terms to zero. Each $\ket{\alpha_{i-1}}$ therefore has the definite charge $q(\alpha_{i-1})$, and so on leftwards along the chain to $\ket{\Psi}$ with charge $Q=q(\alpha_0)$.

Following this reasoning, it can be shown that a state $\ket{\Psi}$ is an eigenstate of $\hat Q$ with eigenvalue $Q$ if and only if it can be written as an abelian-symmetric MPS whose left boundary bond carries the charge $q(\alpha_0)=Q$~\cite{mcculloch2007density,schollwock2011density}. Consequently, if the Hamiltonian posesses the symmetry $[H,\hat{Q}]=0$, then its ground state can be chosen to be an eigenstate of $\hat Q$ and thus has an abelian-symmetric MPS representation with the appropriate left boundary charge, making it a suitable symmetry-constrained ansatz for DMRG. Sorting the bond indices by charge sector then makes every tensor $M_i^{\sigma_i}$ block-sparse, enabling efficient storage and manipulation of the MPS since only the non-zero blocks need to be stored and contracted; the cost is set by the block sizes rather than by the bond dimension (which is their sum). As shown in Ref.~\cite{rupprecht2026faster}, this can also be exploited to reduce state-preparation costs.

We have chosen the less common right-to-left convention for charge accumulation~\cite{mcculloch2007density}. For abelian-symmetric MPS this is a simple relabelling, since the left block at bond $i$ carries the charge $Q-q(\alpha_i)$, but for total spin, whose fusion is not additive, it is a real choice that alters the structure of the tensors themselves. The state-preparation circuit requires the isometric gauge and the charge fusion to point in opposite directions, as explained in Section~\ref{sec:cg-step}. We revisit this construction for spin-adapted MPS in Section~\ref{sec:su2-mps}.

\subsection{Tensor networks with $SU(2)$ symmetry}\label{sec:tns-su2-symmetry}
To exploit the symmetries of chemistry Hamiltonians, we first review how spin-adapted states are structured, starting with spin addition and the representation theory of $\SU2$. The Hamiltonian~\eqref{eqn:chemistry-hamiltonian} commutes with $\hat{S}^2$, but $\hat{S}^2$ is not a charge of the form required by Definition~\ref{def:symmetric-mps}, since it cannot be written as a sum of single-site terms:
\begin{equation}
    \hat S^2 = \sum_{i} \hat{\bm{s}}_i^2 + \sum_{i\neq j} \hat{\bm{s}}_i\cdot\hat{\bm{s}}_j.
\end{equation}
While $\SU2$ invariance does imply the weaker $\hat{S}_z$ symmetry (as used for state preparation in Ref.~\cite{rupprecht2026faster}), $\hat{S}_x$, $\hat{S}_y$ and $\hat{S}_z$ are non-commuting, so exploiting the full compression of the symmetry requires further structure. A \emph{multiplet} of spin $S$ is the $(2S+1)$-dimensional space spanned by the states $\ket{S, M}$, $M = -S, \dots, S$, which rotate into each other under spin rotations. An abelian-symmetric method counts the states of a multiplet individually by their $M$ values. Since the Hamiltonian is invariant under spin rotations, each multiplet can instead be counted once, which gives a $(2S+1)$-fold reduction in the bond dimension of each spin-$S$ sector and a commensurate reduction in the state-preparation cost.

\subsubsection{Spin addition and representation of $\SU2$}
Abelian charges combine with the additive fusion rule~\eqref{eq:charge-cons}, whereas spins combine according to the rules for the coupling of angular momenta~\cite{Sakurai_Napolitano_2017}, which must be respected at each bond.
The total spin operators $\hat S_\alpha = \sum_{i=1}^d \hat s_{\alpha,i}$ commute with $\hat N$, so the Fock space $\mathcal{H}$ of the system of $d$ orbitals decomposes into irreducible representations (irreps) of $\U1\times\SU2$,
\begin{equation}\label{eq:irrep-decomposition}
    \mathcal H = \bigoplus_{N,S} \mathbb{C}^{n_{N,S}} \otimes V_S ,
\end{equation}
where $V_S$ is the $(2S+1)$-dimensional spin-$S$ irrep, spanned by $\ket{S, M}$ with $M = -S,\dots,S$, and $n_{N,S}$ is its \emph{multiplicity}, i.e. the number of times it occurs with particle number $N$.
The irrep decomposition of Equation~\ref{eq:irrep-decomposition} also provides a convenient basis of $\mathcal{H}$, the multiplet basis
\begin{equation}\label{eq:multiplet-basis}
    \ket{a, S, M} = \ket{a}\otimes\ket{S, M},
\end{equation}
where $a=1,\dots,n_{N,S}$ is a \emph{multiplicity label}, which indexes $\mathbb{C}^{n_{N,S}}$ and distinguishes the multiplets of the same particle number and spin (as well as carrying $N$), and the states $\ket{S,M}$ form a \emph{standard basis} of $V_S$: eigenstates of $\hat S_z$ related by the ladder operators with non-negative coefficients
\begin{equation}\label{eq:standard-basis}
    \hat S_\pm \ket{S, M} = \sqrt{(S \mp M)(S \pm M + 1)} \ket{S, M \pm 1}.
\end{equation}
The multiplet basis~\eqref{eq:multiplet-basis} is a key step towards constructing spin-adapted states, since spin rotations act only on the second factor $\ket{S,M}$ (mixing the $2S+1$ states of the multiplet among themselves without changing $a$), allowing all symmetry-constrained components to be factored out and only the multiplicity component $\ket{a}$ to carry non-trivial information about the state. The multiplicity label $a$ is the analogue of the index within a charge sector in the abelian case; the additional structure in $V_S$ captures the non-abelian nature of total-spin symmetry (for an abelian symmetry, this factor would be trivial).

We now consider the coupling of two distinguishable spin systems on Hilbert spaces $\mathcal{H}_1$ and $\mathcal{H}_2$: the composite Hilbert space is the tensor product $\mathcal{H}_1\otimes \mathcal{H}_2$, and the spin operators $\hat{\bm S}_1$, $\hat{\bm S}_2$ each act on their own subsystem only, such that the total spin is $\hat{\bm S}=\hat{\bm S}_1+\hat{\bm S}_2$. Electrons are indistinguishable, and we show in Section~\ref{sec:fermions} that blocks of electronic orbitals can nonetheless be treated in this way. The multiplicity factors combine trivially, and the $V_S$ irreps combine according to the Clebsch-Gordan series
\begin{equation}\label{eq:cg-series}
    V_{S_1} \otimes V_{S_2} \cong \bigoplus_{S = \lvert S_1 - S_2 \rvert}^{S_1 + S_2} V_S ,
\end{equation}
in which each total spin occurs at most once; the spin projections still add with an abelian fusion rule $M=M_1+M_2$, but total spin composes according to the triangular condition
\begin{equation}
S\in\{|S_1-S_2|, |S_1-S_2|+1, \dots, S_1+S_2\},\label{eqn:triangular_condition}
\end{equation}
and the coupled states are linear combinations of the product states
\begin{equation}\label{eq:two-block-coupling}
\begin{aligned}
    \ket{a, b; S, M} &= \sum_{M_1} \cg{S_1}{M_1}{S_2}{M-M_1}{S}{M} \\
    &\quad\times\ket{a, S_1, M_1}\otimes\ket{b, S_2, M-M_1},
\end{aligned}
\end{equation}
where $a$ and $b$ are multiplicity labels for the two subsystems of spin $S_1$ and $S_2$ respectively, the sum over $M_2$ has been eliminated using $M_2 = M - M_1$ (see Definition~\ref{def:cg}), and the sum is weighted by the Clebsch-Gordan coefficients $\cg{S_1}{M_1}{S_2}{M_2}{S}{M}$, defined as follows.
\begin{definition}[Clebsch-Gordan coefficients]\label{def:cg}
For spins $S_1, S_2$ and a total spin $S$ in the Clebsch-Gordan series~\eqref{eq:cg-series}, let $\ket{S, M}$, $M=-S,\dots,S$, be a standard basis of $V_S$ \eqref{eq:standard-basis}. The Clebsch-Gordan coefficients are its components in the product basis,
\begin{equation}\label{eq:cg-def}
    \cg{S_1}{M_1}{S_2}{M_2}{S}{M} \equiv \left(\bra{S_1, M_1} \otimes \bra{S_2, M_2}\right) \ket{S, M},
\end{equation}
which vanish unless $M_1 + M_2 = M$. Equation~\ref{eq:standard-basis} fixes this basis up to one phase for each $S$, which we choose by the Condon-Shortley convention~\cite{condon1935theory}, $\cg{S_1}{S_1}{S_2}{S - S_1}{S}{S} > 0$. All coefficients are then real. The ordering is significant: exchanging the two factors multiplies the coefficient by $(-1)^{S_1 + S_2 - S}$; to keep track of this we write the Clebsch-Gordan coefficients in graphical notation as
\begin{equation}\label{eq:diag-cg-node}
\cg{S_1}{M_1}{S_2}{M_2}{S}{M}
=
\begin{tikzpicture}[
  % the vertex sits on the math axis, so `=' and the equation number line up with it
  baseline={([yshift=-.5ex]C.center)},
  cgleg/.style={line width=1},
  cglabel/.style={font=\scriptsize,inner sep=3pt},
  cgnote/.style={font=\tiny\upshape,text=black!60,inner sep=3.5pt},
  % an arrowhead centred on the middle of a leg, whatever its length: pic[cgmid]{cgtip}
  cgtip/.pic={\draw[line width=1,-{Stealth[length=7pt,width=4pt]}] (-3.5pt,0) -- (3.5pt,0);},
  cgmid/.style={pos=0.5,sloped,allow upside down}]
\Vertex[x=0,y=0,label={$\mathcal{C}$},fontsize=\tiny,position=center,size=0.34,color=gray!30]{C}
% first slot: enters from the right
\draw[cgleg] (1.4,0) -- pic[cgmid]{cgtip}
  node[cglabel,midway,above]{$S_1M_1$} node[cgnote,midway,below]{first input} (C);
% second slot: enters from below; its label and note are stacked under the end of the leg
\draw[cgleg] (0,-0.7) -- pic[cgmid]{cgtip} (C);
\node[cglabel,below,inner sep=2pt] (S2) at (0,-0.7) {$S_2M_2$};
\node[cgnote,below,inner sep=0pt] at (S2.south) {second input};
% coupled spin: leaves to the left
\draw[cgleg] (C) -- pic[cgmid]{cgtip}
  node[cglabel,midway,above]{$S\,M$} node[cgnote,midway,below]{fused output} (-1.4,0);
\end{tikzpicture},
\end{equation}
where arrowed lines carry spins with their projections, and the arrows denote the direction of spin fusion.
\end{definition}
As noted above, we have chosen a right-to-left direction of spin fusion, from the vacuum on the right to the total spin on the left, opposite to the more common left-to-right convention~\cite{paldus1974group,shepard2005general,weichselbaum2012non,keller2016spin}, for reasons that will become apparent in Section~\ref{sec:cg-step}.

\subsubsection{Structural reduction via the Wigner-Eckart theorem} 
The coupling in Equation~\ref{eq:two-block-coupling} acts only on the spin projections --- the multiplicity labels $a$ and $b$ are the same on both sides of the equation, and the Clebsch-Gordan coefficients do not depend on them. It is therefore desirable to try and separate the true physics from the geometric constraints imposed by the spin symmetry --- this is accomplished by the Wigner-Eckart theorem. Each pair of multiplets across the coupling therefore contributes exactly one multiplet of every possible total spin $S$ satisfying the triangular condition $|S_1-S_2|\leq S\leq S_1+S_2$ of the Clebsch-Gordan series~\eqref{eq:cg-series}.
This forms a complete basis of the composite system, accounting for all $(2S_1+1)(2S_2+1)$ product states, which is a multiplet basis of the form in Equation~\ref{eq:multiplet-basis} since each multiplet of total spin $S$ is uniquely identified by the pair of multiplets ($a$ of spin $S_1$ and $b$ of spin $S_2$) that coupled to form it. The Wigner-Eckart theorem shows that the coefficients of any state of definite total spin inherit this structure:
\begin{proposition}[Wigner-Eckart theorem, state form]\label{prop:wigner-eckart-theorem}
(i) Let $\ket{\Psi_{S M}}\in\mathcal{H}$ be a state of the composite system that is an eigenvector of total spin $\hat{\bm S}^2$ and spin projection $\hat S_z$ with eigenvalues $S(S+1)$ and $M$ respectively, expressed in the multiplet basis of the two subsystems as
\begin{equation}
\begin{aligned}
\ket{\Psi_{S M}} &= \sum_{\substack{a, S_1, M_1\\ b, S_2, M_2}} C^{SM}_{(a, S_1, M_1),(b, S_2, M_2)} \\
&\quad\times\ket{a, S_1, M_1}\otimes\ket{b, S_2, M_2},
\end{aligned}
\end{equation}
where the sum runs over all multiplets $(a, S_1)$ of the first subsystem and $(b, S_2)$ of the second, and over their projections $M_i=-S_i,\dots,S_i$.
Then the coefficients $C^{SM}_{(a, S_1, M_1),(b, S_2, M_2)}$ factorise as
\begin{equation}\label{eq:reduced-coefficients}
    C^{SM}_{(a, S_1, M_1),(b, S_2, M_2)} = \cg{S_1}{M_1}{S_2}{M_2}{S}{M} \bm{C}^{SM}_{(a, S_1),(b, S_2)},
\end{equation}
for a \emph{reduced coefficient} $\bm{C}^{SM}_{(a, S_1),(b, S_2)}$ that depends on neither $M_1$ nor $M_2$, where the Clebsch-Gordan coefficient $\cg{S_1}{M_1}{S_2}{M_2}{S}{M}$ vanishes wherever $S$ violates the triangular condition $|S_1-S_2|\leq S\leq S_1+S_2$.

(ii) Let $\{\ket{\Psi_{SM}}\}$, $M=-S,\dots,S$, be a set of $2S+1$ such states that form a standard basis of a multiplet with the phase convention of Equation~\eqref{eq:standard-basis}. Then their reduced coefficients coincide, as $\bm{C}^{SM}_{(a, S_1),(b, S_2)} = \bm{C}^{S}_{(a, S_1),(b, S_2)}$ for every $M$.
Graphically, (i) and (ii) together can be written as
\tikzset{
  cgleg/.style={line width=1},
  cglabel/.style={font=\scriptsize,inner sep=2pt},
  % an arrowhead centred on the middle of a leg, whatever its length: pic[cgmid]{cgtip}
  cgtip/.pic={\draw[line width=1,-{Stealth[length=7pt,width=4pt]}] (-3.5pt,0) -- (3.5pt,0);},
  cgmid/.style={pos=0.5,sloped,allow upside down}}
\begin{equation}\label{eq:diag-wigner-eckart}
% left: the coefficient tensor, one leg per subsystem and one for the state's own label
\begin{tikzpicture}[baseline={([yshift=-.5ex]C.center)}]
\Vertex[x=0,y=0,label={$C$},position=center,size=0.5,color=blue!30]{C}
\draw[cgleg] (C) -- (-0.65,0) node[cglabel,left]{$S\,M$};
\draw[cgleg] (C) -- (0.65,0) node[cglabel,right]{$a\,S_1M_1$};
\draw[cgleg] (C) -- (0,-0.52) node[cglabel,below]{$b\,S_2M_2$};
\end{tikzpicture}
\;=\;
% right: reduced coefficient on the plain (multiplicity) lines, Clebsch-Gordan node on the
% arrowed (spin) lines; the math axis is halfway between the two rows
\begin{tikzpicture}[baseline={([yshift=-.5ex]0,0.275)}]
\Vertex[x=0,y=0.55,label={$\bm C$},position=center,size=0.5,color=red!30]{R}
\Vertex[x=0.8,y=0,label={$\mathcal{C}$},fontsize=\tiny,position=center,size=0.34,color=gray!30]{G}
% spin lines: first slot from the right, second from below, coupled spin out to the left
\draw[cgleg] (1.5,0) node[cglabel,right]{$S_1M_1$} -- pic[cgmid]{cgtip} (G);
\draw[cgleg] (0.8,-0.6) -- pic[cgmid]{cgtip} (G);
\node[cglabel,right,inner sep=4pt] at (0.8,-0.44) {$S_2M_2$};
\draw[cgleg] (G) -- pic[cgmid,pos=0.75]{cgtip} (-0.65,0) node[cglabel,left]{$S\,M$};
% multiplicity lines
\draw[cgleg] (R) -- (-0.65,0.55) node[cglabel,left]{$S$};
\draw[cgleg] (R) -- (1.5,0.55) node[cglabel,right]{$a\,S_1$};
% the b line passes in front of the spin line: a gap in the spin line, then the b line
\draw[white,line width=3.2pt] (0,-0.1) -- (0,0.1);
\draw[cgleg] (R) -- (0,-0.6);
\node[cglabel,left,inner sep=3pt] at (0,-0.44) {$b\,S_2$};
\end{tikzpicture},
\end{equation}
in which the separation of spin projection from the reduced coefficients is visually obvious.
Throughout this work, a spin label that appears on both lines of a pair, i.e. $S_2$ on the lines $b\,S_2$ and $S_2M_2$, is a single index assuming the same value: the pair of lines together carries the one index $(S_2,b,M_2)$ of the subsystem, and a sum over it runs over $S_2$ once.
\end{proposition}
Part (i) concerns a single state of the composite system and is all that the target state requires at any one bipartition, since the eigenstates of the Hamiltonian in Equation~\ref{eqn:chemistry-hamiltonian} can be chosen to have definite $S$ and $M$. It shows that all dependence on the subsystems' spin projections $M_1$ and $M_2$ separates into the Clebsch-Gordan coefficients, which depend only on the representation theory of $\SU2$ and not on the state.
Part (ii) concerns a set of $2S+1$ states, one for each projection $M$. We need it because an MPS describes more than just the contracted target state: the MPS is divided over $d-1$ bonds, at each of which it also defines a basis for the block of orbitals to the left or right of that bond (see Section~\ref{sec:su2-mps}). These block states also have definite spin and therefore come in multiplets of $2S+1$ states. Each of them is a state of a composite system in its own right (i.e. a smaller contiguous block of orbitals being coupled to one more adjacent orbital), so part (i) gives each block state a reduced coefficient $\bm{C}^{SM}$ of its own. Part (ii) shows that a single reduced coefficient $\bm{C}^{S}$ is sufficient to describe the whole multiplet with no $M$-dependence. The tensors of the MPS therefore need one entry per multiplet rather than one per state, and the projection $M$ disappears from their bond indices. This reduction is why adaptation to total-spin $\SU2$ symmetry compresses the state further than the abelian $\U1$ symmetry of the spin projection, and it is the source of the algorithmic improvements reported in this work.

We note that Proposition~\ref{prop:wigner-eckart-theorem} is the \emph{state form} of the Wigner-Eckart theorem; readers familiar with representation theory of $\SU2$ may be more familiar with its operator form, an analogous statement on factorisation of matrix elements of spherical tensor operators~\cite{keller2016spin}. These two forms are essentially equivalent~\cite{sharma2012spin}, but the state form is simpler and pedagogically more transparent for this paper's purposes; the operator form is only required when performing the actual contractions of spin-adapted DMRG~\cite{sharma2012spin,keller2016spin}, whereas we are supplied with the resulting spin-adapted MPS and seek to load them into a quantum computer.

\subsection{Spin-adapted matrix product states}
Proposition~\ref{prop:wigner-eckart-theorem} is a statement on composing the spin states of two subsystems into a total spin state, whereas an MPS has a bond between every pair of adjacent orbitals. Here, we review how applying Proposition~\ref{prop:wigner-eckart-theorem}(ii) at each bond of an MPS gives a non-abelian $\SU2$-adapted version of the symmetric MPS of Section~\ref{sec:abelian-warmup}, whose indices label multiplets rather than individual states and which therefore exploits the full symmetry of the Hamiltonian~\eqref{eqn:chemistry-hamiltonian} to compress the state representation.

\subsubsection{Fermionic conventions}\label{sec:fermions}
Electrons are spin-$\tfrac{1}{2}$ fermions, so each of the $2d$ spin-orbitals $(i,\mu)$ of Section~\ref{sec:problem-setup} holds at most one electron, and the creation and annihilation operators of the spin-orbitals obey the standard anticommutation relations $\{a_{i\mu},a_{j\nu}\}=0$ and $\{a_{i\mu},a^\dagger_{j\nu}\}=\delta_{ij}\delta_{\mu\nu}$.
Since fermionic operators on different spin-orbitals anticommute, the sign of a state depends on the order in which its creation operators are written, thus we fix this by using occupation-number states for occupation numbers $\vec{n}\equiv(n_{1\uparrow},n_{1\downarrow},\cdots, n_{d\uparrow},n_{d\downarrow})$ with the creation operators in increasing order of the spin-orbitals,
\begin{equation}\label{eq:block-product}
    \ket{\vec{n}} \equiv (a^\dagger_{1\uparrow})^{n_{1\uparrow}}(a^\dagger_{1\downarrow})^{n_{1\downarrow}}\cdots(a^\dagger_{d\uparrow})^{n_{d\uparrow}}(a^\dagger_{d\downarrow})^{n_{d\downarrow}}\ket{\bm{0}},
\end{equation}
which matches the Jordan-Wigner order chosen in Section~\ref{sec:problem-setup}. The Fock space of a single spatial orbital (the union of the two spin-orbitals at a site) is spanned by the standard occupation basis $\{\ket{\emptyset}, \ket{\uparrow}, \ket{\downarrow}, \ket{\uparrow\downarrow}\}$, where $\ket{\emptyset}=\ket{00}$ in the computational basis and $\ket{\uparrow}\equiv a^\dagger_{\uparrow}\ket{\emptyset}$, $\ket{\downarrow}\equiv a^\dagger_{\downarrow}\ket{\emptyset}$, and $\ket{\uparrow\downarrow}=a^\dagger_{\uparrow}a^\dagger_{\downarrow}\ket{\emptyset}$. We write the number operator as $\hat n_{i\mu}\equiv a^\dagger_{i\mu}a_{i\mu}$.

A spin-adapted MPS instead uses the multiplet basis of Equation~\ref{eq:multiplet-basis} for each orbital. With the spin operators of orbital $i$ given by $\hat s_{z,i}=\tfrac12(\hat n_{i\uparrow}-\hat n_{i\downarrow})$, $\hat s_{+,i}=a^\dagger_{i\uparrow}a_{i\downarrow}$ and $\hat s_{-,i}=a^\dagger_{i\downarrow}a_{i\uparrow}$, the singly occupied states are mapped to each other by the spin operators, whereas all spin operators annihilate the empty and doubly occupied states, which gives three multiplets labelled by $q=(n,s)$ for occupation $n$ and spin $s$~\cite{keller2016spin}:
\begin{equation}\label{eq:local-multiplets}
\begin{aligned}
    (0,0):&\quad \ket{\emptyset}, \\
    (1,\tfrac12):&\quad \ket{\uparrow}\,\text{or}\, \ket{\downarrow}, \\
    (2,0):&\quad \ket{\uparrow\downarrow},
\end{aligned}
\end{equation}
i.e. a doublet for the singly occupied states and two singlets for the empty and fully occupied states.
In multiplet form, the local basis vectors of the orbital are written as $\ket{q,m}$ for the state of projection $m$ in the multiplet $q$ (with no multiplicity label needed since each multiplet occurs once).

An MPS builds up a state by successively coupling one orbital at a time to the existing block of orbitals, so to apply the results of Section~\ref{sec:tns-su2-symmetry} we must show that a block of orbitals behaves like the distinguishable subsystems of Equation~\ref{eq:two-block-coupling}.
This follows from the ordering of Equation~\ref{eq:block-product} and the matching Jordan-Wigner order. In the Jordan-Wigner encoding the creation operator of a mode carries a string of $Z$ operators on the modes below it; in Equation~\ref{eq:block-product} each creation operator acts before those of the modes below it, finds those modes empty, and its string contributes $+1$. The state $\ket{\vec n}$ is therefore encoded as the bit string $\vec n$, and for occupation-number states $\ket{x}$ and $\ket{y}$ of two adjacent blocks, the occupation-number state with both sets of occupations is the tensor product $\ket{x}\otimes\ket{y}$ of their qubit states.
Since the spin operators of a block are quadratic in fermionic operators of its own orbitals only, they commute with fermionic operators of other blocks and act on their own factor of this product only. Proposition~\ref{prop:wigner-eckart-theorem} therefore applies at every bond.

\subsubsection{The spin-adapted MPS}\label{sec:su2-mps}
With these tools in hand, we can now review how a spin-adapted MPS is constructed in the same manner as an abelian-symmetric MPS was in Section~\ref{sec:abelian-warmup}: sweeping right to left, with each site tensor attaching one more orbital to the right block. Three key differences arise when generalising this to spin-adapted states. First, every index labels a multiplet rather than a state thanks to the structural reduction of the Wigner-Eckart theorem (Proposition~\ref{prop:wigner-eckart-theorem}). Second, we write our system in the multiplet basis rather than the occupation-number basis, so the physical index $\sigma_i$ is replaced by a local multiplet $q_i$ (per Equation~\ref{eq:local-multiplets}), and the bond index $\alpha_i$ is replaced by a multiplet of the right block of orbitals. Third, the basis states of the right block of orbitals and the new orbital are combined not by a simple tensor product, as in the abelian case~\eqref{eq:abelian-recursion}, but by a Clebsch-Gordan coupling to a definite total spin (per Equation~\ref{eq:two-block-coupling}).

The spin-adapted MPS is constructed in terms of the tools of Section~\ref{sec:tns-su2-symmetry}. The role played by the abelian charge $q(\alpha_i)$ in Section~\ref{sec:abelian-warmup} is now played by a pair of quantum numbers: the states of the block of orbitals to the right of bond $i$ come in multiplets, each labelled by a particle number $N_i$ and total spin $S_i$, and these define a \emph{charge sector} $\Gamma_i=(N_i,S_i)$ (analogous to how the dense blocks of a block-sparse tensor in an abelian-symmetric MPS are labelled by their charge). A bond index is defined by a pair $(\Gamma_i,\alpha_i)$, where $\alpha_i$ is a multiplicity index that distinguishes the multiplets of one sector; a single value of the bond index stands for all $2S_i+1$ states of that multiplet, i.e. $\{\ket{\Gamma_i\alpha_i;M}\}_{M=-S_i,\dots,S_i}$.

In analogy to Definition~\ref{def:symmetric-mps}, a spin-adapted MPS can be defined following Refs.~\cite{mcculloch2007density,sharma2012spin,keller2016spin}:
\begin{definition}[Spin-adapted MPS]\label{def:su2-mps}
A spin-adapted MPS is a multiplet of $2S+1$ symmetry-equivalent states $\ket{\Psi_{SM}}$, $M=-S,\dots,S$, with total spin $S$ and projection $M$, whose amplitudes $\psi_{SM}^{\vec{q},\vec{m}} \equiv \braket{q_1m_1,\dots,q_dm_d}{\Psi_{SM}}$ in the local multiplet basis~\eqref{eq:multiplet-basis} may be written as
\begin{equation}\label{eq:su2-mps}
    \psi_{SM}^{\vec{q},\vec{m}}=\sum_{\{\Gamma_i\alpha_iM_i\}}\ \prod_{i=1}^{d} (R_i^{q_i})_{(\Gamma_{i-1}\alpha_{i-1}),(\Gamma_i\alpha_i)}\; \cg{S_i}{M_i}{s_i}{m_i}{S_{i-1}}{M_{i-1}} ,
\end{equation}
where the \emph{reduced tensors} $R_1,\dots,R_d$ are the same for every $M$ and have entries indexed by charge sectors and multiplicity labels.
These entries are the reduced matrix elements of Refs.~\cite{mcculloch2007density,sharma2012spin,keller2016spin}, and $R_i$ collects the degeneracy tensors of Ref.~\cite{singh2012tensor}, one for each combination of sectors.
The right boundary bond is defined as carrying only the trivial vacuum multiplet $\Gamma_d=(0,0)$, and the left boundary bond carries a single multiplet of the target sector, $\Gamma_0=(N,S)$, with the projection of the state, $M_0=M$.
In addition, writing $X_{\Gamma\Gamma'}$ for the block of a matrix $X$ between the sectors $\Gamma$ and $\Gamma'$, a matrix in the multiplicity indices, every reduced tensor obeys the \emph{selection rule}
\begin{equation}\label{eq:su2-selection}
\begin{aligned}
    &(R_i^{q_i})_{\Gamma_{i-1}\Gamma_i} = 0 \ \text{ unless} \\
    &\qquad N_{i-1}=N_i+n_i \\
    &\qquad \text{and }\ S_{i-1}\in\{S_i+s_i,\ \lvert S_i - s_i\rvert\} .
\end{aligned}
\end{equation}
\end{definition}
Definition~\ref{def:su2-mps} is more intuitive in graphical notation, where the amplitudes may be written as
\begin{equation}\label{eq:diag-su2-chain}
    \psi_{SM}^{\vec{q},\vec{m}}=\begin{tikzpicture}[baseline={([yshift=-.5ex]0,0.275)},
  cgleg/.style={line width=1},
  cglabel/.style={font=\scriptsize,inner sep=2pt},
  cgdots/.style={font=\scriptsize,fill=white,inner sep=1.5pt},
  cgtip/.pic={\draw[line width=1,-{Stealth[length=7pt,width=4pt]}] (-3.5pt,0) -- (3.5pt,0);},
  cgmid/.style={pos=0.5,sloped,allow upside down}]
% upper row: reduced tensors; lower row: Clebsch-Gordan nodes
\Vertex[x=0,y=0.55,label={$R_1$},position=center,size=0.5,color=red!30]{R1}
\Vertex[x=1.85,y=0.55,label={$R_2$},position=center,size=0.5,color=red!30]{R2}
\Vertex[x=3.7,y=0.55,label={$R_d$},position=center,size=0.5,color=red!30]{Rd}
\Vertex[x=0.55,y=0,label={$\mathcal{C}$},fontsize=\tiny,position=center,size=0.34,color=gray!30]{C1}
\Vertex[x=2.4,y=0,label={$\mathcal{C}$},fontsize=\tiny,position=center,size=0.34,color=gray!30]{C2}
\Vertex[x=4.25,y=0,label={$\mathcal{C}$},fontsize=\tiny,position=center,size=0.34,color=gray!30]{Cd}
% spin lines, from the vacuum on the right to the total spin on the left; each arrowhead sits
% just beyond the point where the line passes behind the q leg of the site
\draw[cgleg] (4.85,0) node[cglabel,right]{$0\,0$} -- pic[cgmid]{cgtip} (Cd);
\draw[cgleg] (Cd) -- pic[cgmid,pos=0.484]{cgtip} node[cgdots,pos=0.781]{$\cdots$} (C2);
\draw[cgleg] (C2) -- pic[cgmid,pos=0.5]{cgtip}
  node[cglabel,pos=0.682,above]{$S_1M_1$} (C1);
\draw[cgleg] (C1) -- pic[cgmid,pos=0.698]{cgtip} (-0.65,0) node[cglabel,left]{$S\,M$};
% orbital spins: second input of each node, from below
\draw[cgleg] (0.55,-0.6) -- pic[cgmid]{cgtip} (C1);
\draw[cgleg] (2.4,-0.6) -- pic[cgmid]{cgtip} (C2);
\draw[cgleg] (4.25,-0.6) -- pic[cgmid]{cgtip} (Cd);
\node[cglabel,right,inner sep=4pt] at (0.55,-0.44) {$s_1m_1$};
\node[cglabel,right,inner sep=4pt] at (2.4,-0.44) {$s_2m_2$};
\node[cglabel,right,inner sep=4pt] at (4.25,-0.44) {$s_dm_d$};
% multiplet lines
\draw[cgleg] (R1) -- (-0.65,0.55) node[cglabel,left]{$N\,S$};
\draw[cgleg] (R1) -- node[cglabel,midway,above]{$\Gamma_1\alpha_1$} (R2);
\draw[cgleg] (R2) -- node[cgdots,pos=0.662]{$\cdots$} (Rd);
\draw[cgleg] (Rd) -- (4.85,0.55) node[cglabel,right]{$0\,0$};
% the q lines pass in front of the spin line: a gap in the spin line, then the q line
\foreach \x/\i in {0/1,1.85/2,3.7/d}{
  \draw[white,line width=3.2pt] (\x,-0.1) -- (\x,0.1);
  \draw[cgleg] (\x,0.3) -- (\x,-0.6);
  \node[cglabel,left,inner sep=3pt] at (\x,-0.44) {$q_{\i}$};
}
\end{tikzpicture}.
\end{equation}
Here, we see that the reduced tensors (red) carry all the information specific to the state, with one value of the bond index per multiplet, and the Clebsch-Gordan coefficients (grey) encode the information fixed by the symmetry, expanding each multiplet into its $2S_i+1$ states. By the Wigner-Eckart theorem, the state-dependent and symmetry-dependent parts factorise completely in Equation~\ref{eq:diag-su2-chain}: no bond lines connect the two structures.

Coupling a singly-occupied orbital to a right block of orbitals can either raise or lower the total spin, according to the triangular condition~\eqref{eqn:triangular_condition} --- the selection rule~\eqref{eq:su2-selection} therefore allows two possible values of the next spin $S_{i-1}$ for the same $q_i$ and $\Gamma_i$. That is, the labels on the right of a site do not uniquely determine the label on its left by a simple abelian fusion rule, preventing one from directly exploiting it with the protocol of Ref.~\cite{rupprecht2026faster}; the primary purpose of our approach is to present a circuit construction to handle this non-abelian branching explicitly and make a commensurate saving on the quantum resources required.

For each $M$ the state $\ket{\Psi_{SM}}$ is also an ordinary MPS in the sense of Definition~\ref{def:mps}, with physical index $(q_im_i)$ and bond index $(\Gamma_i\alpha_iM_i)$. Its site tensors, which we denote $T_i$, are the factors of the product in Equation~\eqref{eq:su2-mps}:
\begin{equation}\label{eq:site-factorisation}
\begin{aligned}
    &\big(T_i^{(q_i m_i)}\big)_{(\Gamma_{i-1}\alpha_{i-1}M_{i-1}),(\Gamma_i\alpha_i M_i)} \\
    &\qquad= (R_i^{q_i})_{(\Gamma_{i-1}\alpha_{i-1}),(\Gamma_i\alpha_i)}\; \cg{S_i}{M_i}{s_i}{m_i}{S_{i-1}}{M_{i-1}}.
\end{aligned}
\end{equation}
Read in the occupation basis, this MPS conserves the particle number and the spin projection, i.e. it is an abelian-symmetric MPS (Definition~\ref{def:symmetric-mps}) and so is block-sparse. We refer to it as the \emph{non-spin-adapted form} of the state~\cite{zhai2023block2,rupprecht2026faster}.

We work throughout in the right-canonical gauge, i.e. with the $T_i$ satisfying Equation~\eqref{eq:right-canonical}. The Clebsch-Gordan coefficients drop out of this condition (see the proof of Lemma~\ref{lemma:mapping}(iii)), leaving an isometric condition on the reduced tensors alone:
\begin{equation}\label{eq:su2-isometry}
    \Big(\sum_{q_i} R_i^{q_i} (R_i^{q_i})^\dagger\Big)_{\Gamma_{i-1}\Gamma_{i-1}} = \mathds 1,
\end{equation}
for every sector $\Gamma_{i-1}$ of bond $i-1$.

As in the abelian case~\eqref{eq:abelian-recursion}, the state is built up from the right, one orbital at a time. At each site, orbital $i$ is first coupled to a multiplet $\ket{\Gamma_i\alpha_i;M}$ of the right block by Equation~\eqref{eq:two-block-coupling} (with the block as the first subsystem and the orbital as the second),
\begin{equation}\label{eq:site-coupling}
\begin{aligned}
    \ket{\Gamma_i\alpha_i, q_i; S_{i-1}, M} = \sum_{m} &\cg{S_i}{M-m}{s_i}{m}{S_{i-1}}{M} \\
    &\times\ket{q_i, m}\otimes\ket{\Gamma_i\alpha_i; M-m},
\end{aligned}
\end{equation}
and the reduced tensor then combines the coupled states,
\begin{equation}\label{eq:su2-states}
\begin{aligned}
    \ket{\Gamma_{i-1}\alpha_{i-1}; M} = &\sum_{q_i,\Gamma_i,\alpha_i} (R_i^{q_i})_{(\Gamma_{i-1}\alpha_{i-1}),(\Gamma_i\alpha_i)} \\
    &\times \ket{\Gamma_i\alpha_i, q_i; S_{i-1}, M},
\end{aligned}
\end{equation}
for $M=-S_{i-1},\dots,S_{i-1}$, starting from the vacuum $\ket{\Gamma_d,1;0}=\ket{\bm{0}}$ and ending with $\ket{\Gamma_0,1;M}=\ket{\Psi_{SM}}$. Here the doublet of Equation~\eqref{eq:local-multiplets} is a standard basis~\eqref{eq:standard-basis}, since $\hat s_-\ket{\uparrow}=+\ket{\downarrow}$.

In the right-canonical gauge, the right-block states then have the following structure.
\begin{proposition}[Right-block states]\label{prop:su2-mps}
In the right-canonical gauge, the right-block states $\ket{\Gamma_i\alpha_i;M}$, $M=-S_i,\dots,S_i$, of every bond $i$ are orthonormal, and for each $(\Gamma_i,\alpha_i)$ they are a standard basis of a multiplet of particle number $N_i$ and spin $S_i$ of the orbitals $i+1,\dots,d$. Every multiplet of definite $N$ and $S$ has a representation of this kind.
\end{proposition}
The orthonormality is the orthogonality condition of spin-adapted DMRG~\cite{mcculloch2007density}, and the existence follows from the decomposition of a symmetric state into multiplets~\cite{weichselbaum2012non,singh2012tensor}.

We write $n^{(i)}_{\Gamma}$ for the number of multiplets of the sector $\Gamma$ on bond $i$. A spin-adapted MPS then has two different notions of bond dimension on bond $i$: the number of multiplets, $\chi_i=\sum_{\Gamma}n^{(i)}_{\Gamma}$, the range of the bond index $(\Gamma_i\alpha_i)$ of the reduced tensors, and the number of states they span, $\chi^{\ast}_i=\sum_{\Gamma}n^{(i)}_{\Gamma}(2S_\Gamma+1)$, the range of the bond index $(\Gamma_i\alpha_i M_i)$ of the non-spin-adapted form. We write $\chi=\max_i\chi_i$ (i.e., the spin-adapted bond dimension $\chi$ is the multiplet count on the bond) and $\chi^{\ast}=\max_i\chi^{\ast}_i$ (i.e., the non-spin-adapted bond dimension $\chi^{\ast}$ is the state count on the bond), and from here on `bond dimension' without qualification means the spin-adapted $\chi$. For a multiplet target of total spin $S>0$, the number of multiplets on a bond can differ between its description by the right block and by the left block; $\chi_i$ always counts the multiplets of the right block. For a singlet the two descriptions coincide.

\subsection{The unitary group approach}\label{sec:uga}

A spin-adapted MPS is built up one orbital at a time and records, on each bond, the particle number and the spin of the block built so far. These are the same labels used to describe spin-adapted basis states in the \emph{unitary group approach} (UGA) of Paldus~\cite{paldus1974group} and its graphical form (GUGA) due to Shavitt~\cite{shavitt1977graph,shavitt1978matrix}. This correspondence is leveraged in Section~\ref{sec:results}; below we introduce the UGA concepts it relies on.

\paragraph{Steps and step vectors.}
Suppose that we want to attach an orbital $i$ to the block of orbitals to the right of bond $i$, which is in the sector $\Gamma_i=(N_i,S_i)$. The selection rule~\eqref{eq:su2-selection} constrains the sector $\Gamma_{i-1}$ of the enlarged block to take at most four values, and we denote these four possible transitions or \emph{steps} by $\delta_i \in\{0,1,2,3\}$~\cite{paldus1974group, shepard2005general}. The added orbital can be empty ($\delta_i = 0$), singly occupied and raising the spin of the block by $\tfrac12$ ($\delta_i = 1$), singly occupied and lowering the spin of the block by $\tfrac12$ ($\delta_i = 2$, which requires $S_i >0$), or doubly occupied ($\delta_i = 3$). We summarise this in Table~\ref{tab:steps}. Each step changes the sector by a fixed increment $f(\delta)=(\Delta N,\Delta S)$ such that
\begin{equation}\label{eq:step-increment}
    \Gamma_{i-1}=\Gamma_i+f(\delta_i).
\end{equation}
Steps $1$ and $2$ are called positively and negatively spin-coupled~\cite{dobrautz2019efficient}. They both have the local multiplet $(1,\tfrac12)$ and differ in what they do to the spin of the block, not in the projection of the orbital's own spin. We write $q_\delta$ for the local multiplet of step $\delta$ and define a \emph{step vector} as the sequence $\bm\delta=(\delta_1,\dots,\delta_d)$ of the steps of all the orbitals.

\begin{table}[t]
\caption{\label{tab:steps}\textbf{The four steps of a site:} the occupation of the orbital, its local multiplet $q_\delta$, and the increment $f(\delta)=(\Delta N,\Delta S)$ of the sector of the block.}
\begin{ruledtabular}
\begin{tabular}{clccc}
$\delta$ & Orbital & $q_\delta$ & $\Delta N$ & $\Delta S$ \\
\hline
0 & empty & $(0,0)$ & $0$ & $0$ \\
1 & single, positively coupled & $(1,\tfrac12)$ & $1$ & $+\tfrac12$ \\
2 & single, negatively coupled & $(1,\tfrac12)$ & $1$ & $-\tfrac12$ \\
3 & double & $(2,0)$ & $2$ & $0$ \\
\end{tabular}
\end{ruledtabular}
\end{table}

The unitary-group literature attaches the orbitals in increasing order~\cite{paldus1974group,shepard2005general,burkat2025quantum}, but since our bonds are labelled by the right block, we attach them in decreasing order: the step $\delta_i$ takes the block of orbitals $i+1,\dots,d$ to the block of orbitals $i,\dots,d$. The constructions below are therefore the standard ones with the order of the orbitals reversed.

\paragraph{Configuration state functions.}
A step vector along which the block spin $S_i$ never becomes negative describes a path of sectors from the vacuum $\Gamma_d=(0,0)$ to a sector $\Gamma_0=(N,S)$. It also describes a multiplet of states that we denote by $\ket{N,S,M;\bm\delta}$ (with $M \in \{-S,\dots,S \}$). Starting from the vacuum and through the coupling of Equation~\eqref{eq:two-block-coupling}, we add the orbitals in the order $d, d-1, \dots, 1$ one by one. Each time we add an orbital $i$, the first subsystem is the block and the orbital, in the local multiplet $q_{\delta_i}$, is the second. The two are coupled to the spin $S_{i-1}$ induced by $\delta_i$. The multiplet is obtained after the addition of the last orbital. These states are \emph{configuration state functions} (CSFs), the spin-adapted counterparts of Slater determinants, in the genealogical coupling scheme. In the language of Section~\ref{sec:su2-mps}, a CSF is a spin-adapted MPS of bond dimension one: it has a single multiplet on every bond, corresponding to the sector of the path at that bond, and a single non-zero entry (equal to one) in every reduced tensor. By Equation~\eqref{eq:su2-mps}, its amplitudes are 
\begin{equation}\label{eq:csf-amplitudes}
\begin{gathered}
    \braket{q_1m_1,\dots,q_dm_d}{N,S,M;\bm\delta} \;=\; \\
    \sum_{\{M_i\}}\ \prod_{i=1}^{d}\, \cg{S_i}{M_i}{s_i}{m_i}{S_{i-1}}{M_{i-1}},
\end{gathered}
\end{equation}
if $q_i=q_{\delta_i}$ at every site, and zero otherwise. The spins $S_0,\dots,S_d$ correspond to the spins along the path fixed by the step vector, and the sum runs over the projections $M_1,\dots,M_{d-1}$ with $M_0=M$, $M_d=0$. For each $M$ the CSFs form an orthonormal basis of the states of $N$ electrons with total spin $S$ and projection $M$~\cite{paldus1974group,burkat2025quantum}. The UGA usually fixes $M=S$; we allow any $M$, as in Ref.~\cite{burkat2025quantum}, since the circuit of Section~\ref{sec:results} prepares any projection.

For example, consider two electrons in two orbitals with total spin $S=0$. Three step vectors can reach the sector $(2,0)$ from the vacuum. The step vector $\bm\delta=(2,1)$ passes through $(1,\tfrac12)$ on the middle bond: orbital~2 raises the spin from $0$ to $\tfrac12$ and orbital~1 lowers it back to $0$. The corresponding CSF couples the doublet of orbital~2, as the first input of the Clebsch-Gordan coefficient, with the doublet of orbital~1, and yields the open-shell singlet
\begin{equation}\label{eq:two-orbital-singlet}
    \ket{2,0,0;(2,1)} = \tfrac{1}{\sqrt2}\left(a^\dagger_{1\downarrow}a^\dagger_{2\uparrow} - a^\dagger_{1\uparrow}a^\dagger_{2\downarrow}\right)\ket{\bm{0}}.
\end{equation}
The two closed-shell configurations are obtained through the step vectors $(3,0)$ and $(0,3)$.

\begin{figure}
    \centering
    \includegraphics[width=\linewidth]{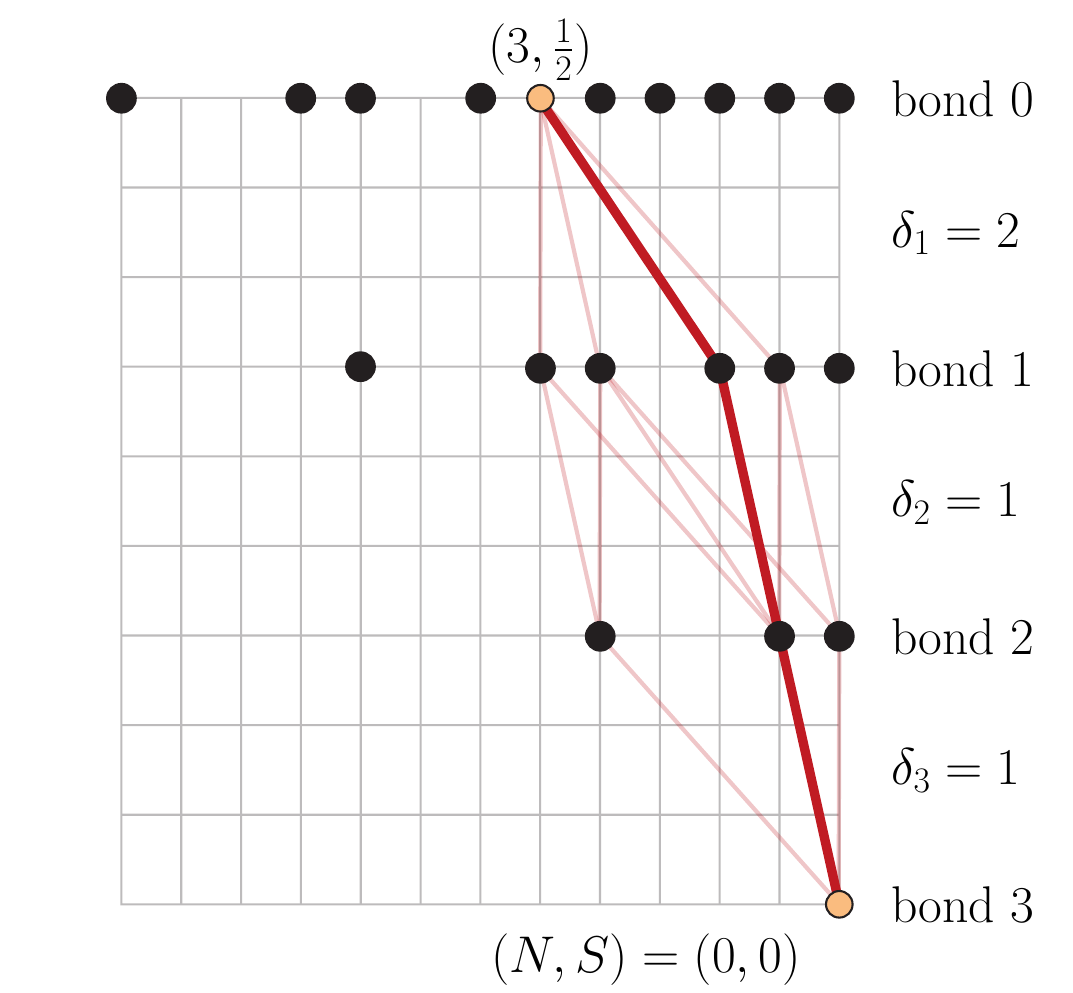}
    \caption{\label{fig:shavitt}\textbf{The Shavitt graph for $d=3$ orbitals.} Each dot is a sector $(N,S)$ of the right block, at a height equal to the number of orbitals in the block and at a distance $2(N-S)$ to the left of the vacuum (one grid column per unit), so that the four steps have four different slopes. The vacuum and the target sector are shown in yellow. The faint lines are the edges of the eight walks from the vacuum to the sector $(3,\tfrac12)$, one for each CSF multiplet of that sector. The thick walk is the step vector $\bm\delta=(2,1,1)$. Because the bonds are labelled by the right block, the first edge belongs to orbital $3$ and the last to orbital $1$.}
\end{figure}

\paragraph{The Shavitt graph.}
The CSFs can be represented compactly with the \emph{Shavitt graph}~\cite{shavitt1977graph,shavitt1978matrix} (see Figure~\ref{fig:shavitt}). Its nodes are the sectors $(N,S)$ of a block of $k$ orbitals (with $k \in \{0,\dots,d\}$), arranged in levels by $k$. An edge joins two nodes on neighbouring levels if they differ by the increment of a step, as summarised in Table~\ref{tab:steps}. A step vector is thus described by a walk on this graph, starting from the vacuum and ending at a target sector. The walks ending at a given node are in one-to-one correspondence with the CSF multiplets of the block in that sector~\cite{shepard2005general,dobrautz2019efficient}, and a single CSF is obtained by choosing a walk together with a projection $M$. The graph is far smaller than the basis it describes. For even $d$, the walks to a singlet with $N=d$ pass through $(d+2)(d+3)(d+4)/24$ nodes in total, while the number of CSFs grows as $4^d/d^2$~\cite{dobrautz2019efficient}. The number of nodes bounds only the number of sectors on each bond, not the bond dimension of a spin-adapted MPS, since each sector can carry many multiplets.

\paragraph{Group-theoretical meaning.}
The steps also have a representation-theoretic interpretation~\cite{burkat2025quantum}. The orbital rotations $\U{d}$ and the spin group $\SU2$ act on the Fock space with commuting actions, and the space decomposes as
\begin{equation}\label{eq:paldus-duality}
    \bigwedge\left(\mathbb C^d\otimes\mathbb C^2\right) \cong \bigoplus_{N,S} W^{\U{d}}_{(N,S)}\otimes V_S ,
\end{equation}
with a single irreducible representation $W^{\U{d}}_{(N,S)}$ of $\U{d}$ for each sector. This splitting holds for any block of orbitals. If $\U{k}$ denotes the rotations of the last $k$ orbitals, its irreducible representations are also labelled by the sectors of that block. When going from $\U{k}$ to $\U{k-1}$ (i.e.\ when removing one orbital from the block), each irreducible representation of $\U{k}$ can be decomposed into irreducible representations of $\U{k-1}$, with exactly one for each allowed step~\cite{burkat2025quantum}. By going from $\U{d}$ down to $\U{1}$, we see that each basis vector of $W^{\U{d}}_{(N,S)}$ is labelled by a succession of steps, i.e.\ a step vector. The basis constructed in such a way is called the Gelfand--Tsetlin basis~\cite{dobrautz2019efficient}. For each $M$, the CSFs are the vectors of this basis tensored with $\ket{S,M}$, up to a phase~\cite{burkat2025quantum}. In particular, the multiplicity label $a$ of Equation~\eqref{eq:multiplet-basis} can be taken to be a step vector.

\paragraph{The quantum Paldus transform.}
Burkat and Fitzpatrick~\cite{burkat2025quantum} introduced a quantum circuit, the \emph{quantum Paldus transform}, that performs the change of basis between CSFs and step vectors. It is an isometry which takes a CSF, stored on the $2d$ qubits of the orbitals in the occupation basis, together with three ancillary registers initialised in $\ket{0}$, to a product state of four registers,
\begin{equation}\label{eq:qpt}
    \ket{0}\ket{0}\ket{0}\otimes\ket{N,S,M;\bm\delta} \;\mapsto\; \ket{N}\ket{S}\ket{M}\ket{\bm\delta}.
\end{equation}
Here $N$, $S$ and $M$ are written in binary on $\mathcal{O}(\log d)$ qubits each, and the step vector occupies the $2d$ qubits of the orbitals (two qubits for each step). The circuit is a cascade of $d$ Clebsch--Gordan transforms, one for each orbital. Each of them takes the occupation of its orbital to a step through rotations controlled by the $S$ and $M$ registers, and then updates the $N$, $S$ and $M$ registers accordingly. In total, the circuit applies $d(d+1)(d+2)/6$ controlled rotations, and Ref.~\cite{burkat2025quantum} reports a fault-tolerant cost of $\mathcal{O}(d^3)$ Toffoli gates. When run in reverse on $\ket{N}\ket{S}\ket{M}\sum_{\bm\delta}c_{\bm\delta}\ket{\bm\delta}$, where every $\bm\delta$ is a valid walk to the sector $(N,S)$, it returns the ancillary registers to $\ket{0}$ and prepares $\sum_{\bm\delta}c_{\bm\delta}\ket{N,S,M;\bm\delta}$, at a cost that does not depend on the state. Reference~\cite{burkat2025quantum} attaches the orbitals in increasing order, and we relate the transform to our order in Section~\ref{sec:cg-step}.

\section{Results}\label{sec:results}
A spin-adapted MPS compresses each multiplet by a factor of $2S+1$ by storing only one amplitude; here we construct a state-preparation circuit whose cost reflects this compression. The protocol of Ref. \cite{rupprecht2026faster} cannot be applied directly to a spin-adapted MPS, because it assumes an additive abelian selection rule~\eqref{eq:charge-cons}, in which the local state determines how the bond charge changes. As outlined in Section~\ref{sec:su2-mps}, spin fusion does not follow such a rule, since a singly-occupied orbital can either raise or lower the total spin of a system it is coupled to. This obstacle can be removed in three stages:
\begin{enumerate}
    \item Relabel: use the conventions of the unitary group approach (Section~\ref{sec:uga}) to label each site by its step (Table~\ref{tab:steps}) rather than its local state. In terms of steps, the sector label is additive, and the reduced tensors become an abelian-symmetric MPS with four local states per site (Lemma~\ref{lemma:mapping}). This is a simple classical relabelling of amplitudes at no additional cost.
    \item Load: the relabelling `abelian-ises' the problem, allowing one to directly apply the block-sparse preparation of Ref.~\cite{rupprecht2026faster} to efficiently load an MPS over step vectors, whose sector labels act like an abelian charge (Corollary~\ref{corollary:loading}) at cost determined by the multiplet count as desired.
    \item Couple: the register now holds unitary-group-approach steps, not occupations. At each site an inverse Clebsch--Gordan transform (a fixed transform determined by the geometry of the symmetry group) turns the step of a singly-occupied orbital into the correct superposition of spin states (Section~\ref{sec:cg-step}), yielding the desired quantum state in the occupation basis. Its cost depends on the number of orbitals but not on the bond dimension, and is negligible compared to that of the data loading.
\end{enumerate}
The second and third stages can be interleaved site by site, reading the spin from the bond register instead of a separate spin register; this saves a little when the bond register stores the sector explicitly, and either way the difference is small (Section~\ref{sec:cost}). The result is a single circuit that prepares the state for any projection $M$ (Theorem~\ref{thm:preparation} and Figure~\ref{fig:site-circuit}), at a cost dominated by the step-vector-MPS loading, which is itself determined by the multiplet count $\chi$ (Section~\ref{sec:cost}) rather than the full bond dimension $\chi^{\ast}>\chi$.

\subsection{A spin-adapted MPS is an abelian-symmetric MPS over step vectors}\label{sec:mapping}
Accordingly, we relabel the physical (multiplet) index of the reduced tensors $q_i$ by the steps of Table~\ref{tab:steps}, in order to map it into a form amenable to state preparation:
\begin{definition}[Step-vector MPS]\label{def:step-mps}
The \emph{step-vector MPS} of a spin-adapted MPS is the MPS on $d$ sites of local dimension $D=4$ whose tensors $A_i^{\delta}$, $\delta\in\{0,1,2,3\}$, the \emph{step-vector tensors}, are
\begin{equation}\label{eq:step-tensor}
    (A_i^{\delta})_{\Gamma_{i-1}\Gamma_i} = (R_i^{q_\delta})_{\Gamma_{i-1}\Gamma_i},
\end{equation}
if $\Gamma_{i-1}=\Gamma_i+f(\delta)$, and $0$ otherwise.
Contracted, the tensors define the state
\begin{equation}\label{eq:step-state}
    \ket{\Phi} = \sum_{\bm\delta} c_{\bm\delta}\,\ket{\delta_1}\ket{\delta_2}\cdots\ket{\delta_d},
    \quad c_{\bm\delta}=A_1^{\delta_1}A_2^{\delta_2}\cdots A_d^{\delta_d}.
\end{equation}
Each $c_{\bm\delta}$ is a scalar since both boundary bonds of the spin-adapted MPS are trivial (Definition~\ref{def:su2-mps}).
\end{definition}
The step-vector tensors contain the same amplitudes as the reduced tensors, relabelled (Lemma~\ref{lemma:mapping}(ii) below). Viewed as matrices for a fixed value of the physical index, each $A_i^{\delta}$ has the same row and column structure as the corresponding $R_i^{q_\delta}$; the condition $\Gamma_{i-1}=\Gamma_i+f(\delta)$ merely sets some of its blocks to zero. This condition only applies for $\delta\in\{1,2\}$ since the selection rule~\eqref{eq:su2-selection} already enforces it for $\delta\in\{0,3\}$, therefore $A_i^0=R_i^{(0,0)}$ and $A_i^3=R_i^{(2,0)}$.
For $\delta=1$ and $\delta=2$ it splits the non-zero blocks of $R_i^{(1,1/2)}$ in two: the positively spin-coupled blocks go to $A_i^1$ and the negatively spin-coupled ones to $A_i^2$. We depict this splitting graphically in Figure~\ref{fig:step-blocks}. The mapping has the following properties:
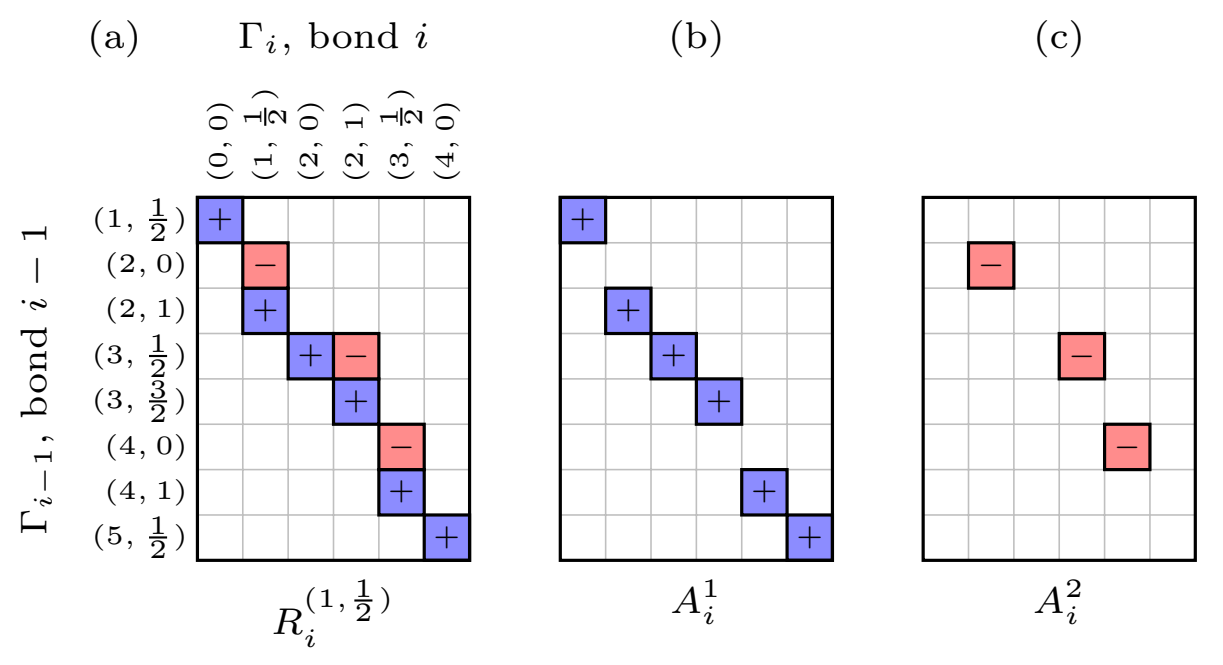
\begin{figure}
\centering
\begin{tikzpicture}[x=1cm,y=1cm,line width=0.6pt,
  blab/.style={font=\scriptsize,inner sep=2pt},
  tlab/.style={font=\tiny,inner sep=1.5pt}]
% One slice of a site tensor per panel. Rows: sectors of bond i-1 (three orbitals to the right),
% columns: sectors of bond i (two orbitals to the right). Cell (r,c) spans
% x in [0.3(c-1), 0.3c] and y in [2.4-0.3r, 2.4-0.3(r-1)], shifted by the panel offset.
% Blocks of a singly occupied orbital, (row, column):
%   positively coupled, Gamma_{i-1} = Gamma_i + (1,+1/2): (1,1) (3,2) (4,3) (5,4) (7,5) (8,6)
%   negatively coupled, Gamma_{i-1} = Gamma_i + (1,-1/2): (2,2) (4,4) (6,5)
\foreach \ox/\pos/\negs/\name in {0/1/1/{$R_i^{(1,\frac12)}$}, 2.4/1/0/{$A_i^{1}$}, 4.8/0/1/{$A_i^{2}$}} {
    \begin{scope}[shift={(\ox,0)}]
    \draw[black!25, line width=0.3pt] (0,0) grid[step=0.3] (1.8,2.4);
    \ifnum\pos=1
        \foreach \r/\c in {1/1, 3/2, 4/3, 5/4, 7/5, 8/6} {
            \filldraw[fill=blue!45, draw=black] ({0.3*(\c-1)},{2.4-0.3*\r}) rectangle ({0.3*\c},{2.7-0.3*\r});
            \node[tlab] at ({0.3*\c-0.15},{2.55-0.3*\r}) {$+$};
        }
    \fi
    \ifnum\negs=1
        \foreach \r/\c in {2/2, 4/4, 6/5} {
            \filldraw[fill=red!45, draw=black] ({0.3*(\c-1)},{2.4-0.3*\r}) rectangle ({0.3*\c},{2.7-0.3*\r});
            \node[tlab] at ({0.3*\c-0.15},{2.55-0.3*\r}) {$-$};
        }
    \fi
    \draw (0,0) rectangle (1.8,2.4);
    \node[blab,below] at (0.9,-0.05) {\name};
    \end{scope}
}
% sector labels, on the first panel only
\foreach \r/\lab in {1/{(1,\frac12)}, 2/{(2,0)}, 3/{(2,1)}, 4/{(3,\frac12)}, 5/{(3,\frac32)}, 6/{(4,0)}, 7/{(4,1)}, 8/{(5,\frac12)}}
    \node[tlab,left] at (0,{2.55-0.3*\r}) {$\lab$};
\foreach \c/\lab in {1/{(0,0)}, 2/{(1,\frac12)}, 3/{(2,0)}, 4/{(2,1)}, 5/{(3,\frac12)}, 6/{(4,0)}}
    \node[tlab,above,rotate=90,anchor=west] at ({0.3*\c-0.15},2.43) {$\lab$};
\node[blab,rotate=90] at (-1.05,1.2) {$\Gamma_{i-1}$, bond $i-1$};
\node[blab] at (0.9,3.45) {$\Gamma_i$, bond $i$};
\node[blab] at (3.3,3.45) {(b)};
\node[blab] at (5.7,3.45) {(c)};
\node[blab] at (-0.55,3.45) {(a)};
\end{tikzpicture}
\caption{\label{fig:step-blocks}\textbf{A graphical depiction of Definition~\ref{def:step-mps} at one site.} Here, we consider a site $i$ of a spin-adapted MPS with two orbitals to its right. Each panel depicts one matrix slice of a tensor $R_i$ or $A_i$, with rows and columns labelled by the sectors of bonds $i-1$ and $i$, so that each square is one block $X_{\Gamma_{i-1}\Gamma_i}$; white squares are zero, blue squares are blocks of amplitudes for positive spin coupling, red squares are blocks of amplitudes for negative spin coupling. \textbf{(a)} The slice $R_i^{(1,1/2)}$ of the reduced tensor, for a singly occupied orbital. \textbf{(b, c)} The step-vector tensors $A_i^1$ and $A_i^2$ keep the positively-coupled and negatively-coupled blocks respectively, with the same entries.}
\end{figure}
\begin{lemma}[Step-vector mapping of a spin-adapted MPS]\label{lemma:mapping}
Let $R_1,\dots,R_d$ be tensors of a spin-adapted MPS (Definition~\ref{def:su2-mps}) representing the multiplet $\{\ket{\Psi_{SM}}\}_{M=-S,\dots,S}$, and let $\ket{\Phi}$ be the state~\eqref{eq:step-state} of its step-vector MPS. Then:
\textbf{(i)} Choosing the additive charge to be the pair $\Gamma=(N,S)$, carried by the bond index $(\Gamma_i\alpha_i)$ as $\Gamma_i$ and by the local state $\ket{\delta}$ as $f(\delta)$, the step-vector MPS (Definition~\ref{def:step-mps}) is an abelian-symmetric MPS (Definition~\ref{def:symmetric-mps}).
\textbf{(ii)} Every entry of a reduced tensor $R_i^{q}$ that the selection rule~\eqref{eq:su2-selection} permits is an entry of exactly one step-vector tensor $A_i^{\delta}$.
\textbf{(iii)} The step-vector MPS is right-canonical, $\sum_\delta A_i^\delta (A_i^\delta)^\dagger=\mathds 1$, if and only if the spin-adapted MPS is.
\textbf{(iv)} For every spin projection $M$, the target is obtained from $\ket{\Phi}$ by replacing each step basis state $\ket{\delta_1}\cdots\ket{\delta_d}$ with the CSF $\ket{N,S,M;\bm\delta}$, keeping the amplitudes:
\begin{equation}\label{eq:csf-expansion}
\begin{aligned}
    \ket{\Phi} &= \sum_{\bm\delta} c_{\bm\delta}\,\ket{\delta_1}\cdots\ket{\delta_d} \\
    \mapsto\quad \ket{\Psi_{SM}} &= \sum_{\bm\delta} c_{\bm\delta}\,\ket{N,S,M;\bm\delta},
\end{aligned}
\end{equation}
with the same amplitudes $c_{\bm\delta}$ for all $2S+1$ projections. Only step vectors that are legal paths (from the vacuum to $(N,S)$, with the spin never negative) have a CSF, and every other step vector has $c_{\bm\delta}=0$.
\end{lemma}

\begin{proof}
\textbf{(i)} The site charge of Definition~\ref{def:symmetric-mps} is induced by $\hat Q_i=\sum_\delta f(\delta)\ket{\delta}\!\bra{\delta}$, which is diagonal in the step basis. By Equation~\eqref{eq:step-tensor}, a non-zero block of $A_i^{\delta}$ has $\Gamma_{i-1}=\Gamma_i+f(\delta)$, which is the selection rule~\eqref{eq:charge-cons} with the bond charge $\Gamma_i$ and the site charge $f(\delta)$. The right boundary bond carries $\Gamma_d=(0,0)$ by Definition~\ref{def:su2-mps}.

\textbf{(ii)} Choose a site $i$ and a block $(R_i^{q_i})_{\Gamma_{i-1}\Gamma_i}$ permitted by the selection rule~\eqref{eq:su2-selection}. We consider each local multiplet in turn. Unoccupied or doubly-occupied orbitals are straightforward: for $q_i=(0,0)$, the selection rule gives $N_{i-1}=N_i$ and $S_{i-1}=S_i$, the increment of step $0$, which is the only step with this local multiplet; likewise $q_i=(2,0)$ gives $N_{i-1}=N_i+2$ and $S_{i-1}=S_i$, the increment of step $3$. In the singly-occupied case $q_i=(1,\tfrac12)$, the selection rule gives $N_{i-1}=N_i+1$ and $S_{i-1}\in\{S_i+\tfrac12,| S_i-\tfrac12|\}$.  For $S_i>0$ these are the two distinct values $S_i\pm\tfrac12$, the increments of steps $\delta=1$ and $\delta=2$, so exactly one of these must match the block.
For $S_i=0$ the two values permitted by the selection rule~\eqref{eq:su2-selection} coincide at $S_{i-1}=\tfrac12$, which only step $\delta=1$ reaches; step $\delta=2$ would lower the spin to $-\tfrac12$.
In every case exactly one step $\delta$ has $q_\delta=q_i$ and $\Gamma_{i-1}=\Gamma_i+f(\delta)$, so by Equation~\eqref{eq:step-tensor} the block belongs to that $A_i^{\delta}$ only.

\textbf{(iii)} We compare the two conditions through a third, which involves the reduced tensors alone: Equation~\eqref{eq:su2-isometry}, for every sector $\Gamma_{i-1}$ of bond $i-1$.
We show below that in each of the two conditions the sum is block-diagonal in the sector of bond $i-1$, and that its diagonal blocks are those of $\sum_{q_i}R_i^{q_i}(R_i^{q_i})^\dagger$ (for the spin-adapted MPS, times the identity on the projections).
Each sum is therefore the identity exactly when Equation~\eqref{eq:su2-isometry} holds for every $\Gamma_{i-1}$, so the two conditions hold or fail together.

\emph{The step-vector MPS.} The contraction of $A_i^\delta$ and its adjoint is computed as
\begin{equation}\label{eq:AAdag-diag}
\tikzset{
  cgleg/.style={line width=1},
  cglabel/.style={font=\scriptsize,inner sep=2pt}}
\begin{aligned}
&\Big(\sum_\delta A_i^\delta (A_i^\delta)^\dagger\Big)_{\Gamma_{i-1}\Gamma'_{i-1}}
=\;
\begin{tikzpicture}[baseline={([yshift=-.5ex]0,0)}]
\colorlet{cgline}{.}
\Vertex[x=0,y=0.45,label={$A_i$},position=center,size=0.5,color=red!30]{A}
\Vertex[x=0,y=-0.45,label={$A_i^\dagger$},position=center,size=0.5,color=red!30]{Ac}
\draw[cgleg] (A) -- (-0.65,0.45) node[cglabel,left]{$\Gamma_{i-1}\alpha_{i-1}$};
\draw[cgleg] (Ac) -- (-0.65,-0.45) node[cglabel,left]{$\Gamma'_{i-1}\alpha'_{i-1}$};
\Edge[lw=1,color=cgline](A)(Ac)
\node[cglabel,left,inner sep=3pt] at (0,0) {$\delta$};
\draw[cgleg] (A) -- (0.6,0.45) to[out=0,in=0,looseness=1.5] (0.6,-0.45) -- (Ac);
\node[cglabel,above] at (0.7,0.45) {$\Gamma_i\alpha_i$};
\end{tikzpicture}
\\[4pt]
&\quad=\;\delta_{\Gamma_{i-1}\Gamma'_{i-1}}\,
\begin{tikzpicture}[baseline={([yshift=-.5ex]0,0)}]
\colorlet{cgline}{.}
\Vertex[x=0,y=0.45,label={$R_i$},position=center,size=0.5,color=red!30]{R}
\Vertex[x=0,y=-0.45,label={$R_i^\dagger$},position=center,size=0.5,color=red!30]{Rc}
\draw[cgleg] (R) -- (-0.65,0.45) node[cglabel,left]{$\Gamma_{i-1}\alpha_{i-1}$};
\draw[cgleg] (Rc) -- (-0.65,-0.45) node[cglabel,left]{$\Gamma'_{i-1}\alpha'_{i-1}$};
\Edge[lw=1,color=cgline](R)(Rc)
\node[cglabel,left,inner sep=3pt] at (0,0) {$q_i$};
\draw[cgleg] (R) -- (0.6,0.45) to[out=0,in=0,looseness=1.5] (0.6,-0.45) -- (Rc);
\node[cglabel,above] at (0.7,0.45) {$\Gamma_i\alpha_i$};
\end{tikzpicture},
\end{aligned}
\end{equation}
where we have applied Equation~\ref{eq:step-tensor} to both halves of each term, which share $\delta$ and $\Gamma_i$, and used (ii) to turn the sum over $\delta$ into one over $q_i$. The condition is therefore Equation~\eqref{eq:su2-isometry}.

\emph{The spin-adapted MPS.} Its condition is Equation~\ref{eq:right-canonical} for the site tensors $T_i$, each drawn as one site of the diagram~\eqref{eq:diag-su2-chain}:
\begin{equation}\label{eq:TTdag-diag}
\tikzset{
  cgleg/.style={line width=1},
  cglabel/.style={font=\scriptsize,inner sep=2pt},
  % an arrowhead centred on the middle of a leg, whatever its length: pic[cgmid]{cgtip}
  cgtip/.pic={\draw[line width=1,-{Stealth[length=7pt,width=4pt]}] (-3.5pt,0) -- (3.5pt,0);},
  cgmid/.style={pos=0.5,sloped,allow upside down}}
\begin{aligned}
&\Big(\sum_{q_i,m_i}T_i^{(q_im_i)}(T_i^{(q_im_i)})^\dagger\Big)_{(\Gamma_{i-1}\alpha_{i-1}M_{i-1}),(\Gamma'_{i-1}\alpha'_{i-1}M'_{i-1})} \\[2pt]
% upper half: T_i, drawn as one site of eq:diag-su2-chain. Lower half: its conjugate, the
% mirror image with the arrows reversed. They are joined on the orbital and on bond i.
&\quad=\;\begin{tikzpicture}[baseline={([yshift=-.5ex]0,0)}]
% tikz-network edges in the colour of the surrounding text, like the plain legs
\colorlet{cgline}{.}
\Vertex[x=0,y=1.0,label={$R_i$},position=center,size=0.5,color=red!30]{R}
\Vertex[x=0,y=-1.0,label={$R_i^\dagger$},position=center,size=0.5,color=red!30]{Rc}
\Vertex[x=0.8,y=0.45,label={$\mathcal{C}$},fontsize=\tiny,position=center,size=0.34,color=gray!30]{C}
\Vertex[x=0.8,y=-0.45,label={$\mathcal{C}$},fontsize=\tiny,position=center,size=0.34,color=gray!30]{Cc}
% spin lines of bond i-1: out of the upper node, into the lower one
\draw[cgleg] (C) -- pic[cgmid,pos=0.75]{cgtip} (-0.65,0.45) node[cglabel,left]{$S_{i-1}M_{i-1}$};
\draw[cgleg] (-0.65,-0.45) node[cglabel,left]{$S'_{i-1}M'_{i-1}$} -- pic[cgmid,pos=0.25]{cgtip} (Cc);
% spin line of bond i (first slot of both nodes) and spin of the orbital (second slot of both)
\draw[cgleg] (Cc) -- (1.5,-0.45) to[out=0,in=0,looseness=1.5] pic[cgmid]{cgtip} (1.5,0.45) -- (C);
\node[cglabel,above] at (1.4,0.45) {$S_iM_i$};
\Edge[lw=1,color=cgline](Cc)(C)
\path (Cc) -- pic[cgmid]{cgtip} (C);
\node[cglabel,right,inner sep=4pt] at (0.8,0) {$s_im_i$};
% multiplicity lines
\draw[cgleg] (R) -- (-0.65,1.0) node[cglabel,left]{$\Gamma_{i-1}\alpha_{i-1}$};
\draw[cgleg] (Rc) -- (-0.65,-1.0) node[cglabel,left]{$\Gamma'_{i-1}\alpha'_{i-1}$};
\draw[cgleg] (R) -- node[cglabel,pos=0.55,above]{$\Gamma_i\alpha_i$} (1.65,1.0)
  to[out=0,in=0] (1.65,-1.0) -- (Rc);
% the q line passes in front of both spin lines: a gap in each spin line, then the q line
\draw[white,line width=3.2pt] (0,0.35) -- (0,0.55);
\draw[white,line width=3.2pt] (0,-0.55) -- (0,-0.35);
\draw[cgleg] (R) -- (Rc);
\node[cglabel,left,inner sep=3pt] at (0,0) {$q_i$};
\end{tikzpicture}
\\[4pt]
% the two Clebsch-Gordan nodes contract to one spin line, a Kronecker delta
&\quad=\;
\begin{tikzpicture}[baseline={([yshift=-.5ex]0,0)}]
\colorlet{cgline}{.}
\Vertex[x=0,y=0.45,label={$R_i$},position=center,size=0.5,color=red!30]{R}
\Vertex[x=0,y=-0.45,label={$R_i^\dagger$},position=center,size=0.5,color=red!30]{Rc}
% multiplicity lines
\draw[cgleg] (R) -- (-0.65,0.45) node[cglabel,left]{$\Gamma_{i-1}\alpha_{i-1}$};
\draw[cgleg] (Rc) -- (-0.65,-0.45) node[cglabel,left]{$\Gamma'_{i-1}\alpha'_{i-1}$};
\Edge[lw=1,color=cgline](R)(Rc)
\node[cglabel,left,inner sep=3pt] at (0,0) {$q_i$};
\draw[cgleg] (R) -- (0.6,0.45) to[out=0,in=0,looseness=1.5] (0.6,-0.45) -- (Rc);
\node[cglabel,above] at (0.7,0.45) {$\Gamma_i\alpha_i$};
\end{tikzpicture}
\;
\underbrace{%
\begin{tikzpicture}[baseline={([yshift=-.5ex]0,0)}]
\draw[cgleg] (0,-0.45) node[cglabel,left]{$S'_{i-1}M'_{i-1}$} -- (0.6,-0.45)
  to[out=0,in=0,looseness=1.5] pic[cgmid]{cgtip} (0.6,0.45)
  -- (0,0.45) node[cglabel,left]{$S_{i-1}M_{i-1}$};
\end{tikzpicture}}_{\delta_{S_{i-1}S'_{i-1}}\,\delta_{M_{i-1}M'_{i-1}}}.
\end{aligned}
\end{equation}
Since $q_i$ and $\Gamma_i$ are shared, the selection rule~\eqref{eq:su2-selection} also gives $N_{i-1}=N'_{i-1}$, so the right-hand side is $\delta_{M_{i-1}M'_{i-1}}$, and the condition is again Equation~\eqref{eq:su2-isometry}.

\textbf{(iv)} Group the terms of Equation~\eqref{eq:su2-mps} by their sequence of sectors $\Gamma_0,\dots,\Gamma_d$. By (ii), each such sequence corresponds to one step vector $\bm\delta$.
For a fixed $\bm\delta$ the reduced-tensor entries depend only on the multiplicity indices and the Clebsch-Gordan coefficients only on the projections, so Equation~\eqref{eq:su2-mps} factorises:
\begin{equation}
\begin{aligned}
    \psi_{SM}^{\vec{q},\vec{m}}
    &= \sum_{\bm\delta}\Big[\sum_{\{\alpha_i\}}\prod_{i=1}^{d}(R_i^{q_i})_{(\Gamma_{i-1}\alpha_{i-1}),(\Gamma_i\alpha_i)}\Big] \\
    &\qquad\times\Big[\sum_{\{M_i\}}\prod_{i=1}^{d}\cg{S_i}{M_i}{s_i}{m_i}{S_{i-1}}{M_{i-1}}\Big] \\
    &= \sum_{\bm\delta} c_{\bm\delta}\braket{q_1m_1,\dots,q_dm_d}{N,S,M;\bm\delta},
\end{aligned}
\end{equation}
where in each term the sectors $\Gamma_i$ are those of the path of $\bm\delta$. The first bracket is $A_1^{\delta_1}\cdots A_d^{\delta_d}=c_{\bm\delta}$ by Equation~\eqref{eq:step-tensor}, and the second is Equation~\eqref{eq:csf-amplitudes}. This yields the second line of Equation~\eqref{eq:csf-expansion}, whose coefficients are those of $\ket{\Phi}$ by Equation~\eqref{eq:step-state}. For illegal paths, the condition of Equation~\eqref{eq:step-tensor} fixes the sectors of $c_{\bm\delta}=A_1^{\delta_1}\cdots A_d^{\delta_d}$ one bond at a time, $\Gamma_{i-1}=\Gamma_i+f(\delta_i)$, from $\Gamma_d=(0,0)$. For this to reach a negative spin or end at a sector other than $(N,S)$, some bond would need a sector it does not actually carry, since no bond has negative spin and bond $0$ carries only $(N,S)$; it follows that $c_{\bm\delta}=0$ in these cases.
\end{proof}
\subsection{Preparing the step-vector MPS}\label{sec:loading}

Lemma~\ref{lemma:mapping} places spin-adapted states within the scope of Ref.~\cite{rupprecht2026faster} by casting them as abelian-symmetric MPS; we now show how to prepare the step-vector MPS efficiently. An MPS is typically prepared sequentially, one site at a time, with a bond register of $\lceil\log_2\chi\rceil$ qubits that carries the entanglement between the sites already prepared and the rest of the chain \cite{schoen2005sequential,fomichev2024initial,berry2025rapid,rupprecht2026faster}. One can save $\lceil\log_2\chi\rceil$ ancillary qubits by letting the bond register `float' along the uninitialised physical-site qubits. We did not find this saving in the literature and prove it in Appendix~\ref{app:floating}.
The bond register is initialised in the state $\ket{\alpha_0}$, and at each site $i=1,\dots,d$ a unitary acts on it and on the register of site $i$, initialised in $\ket{0}$, as
\begin{equation}\label{eq:site-unitary}
    \ket{\alpha_{i-1}}\ket{0} \mapsto \sum_{\sigma_i,\alpha_i} (M_i^{\sigma_i})_{\alpha_{i-1}\alpha_i} \ket{\alpha_i}\ket{\sigma_i},
\end{equation}
which is an isometry by the right-canonical condition~\eqref{eq:right-canonical} (since the columns are orthonormal); the unitary therefore need only reproduce these columns, and the isometry synthesis can be decomposed into smaller unitary synthesis problems accordingly~\cite{berry2025rapid}. For an abelian-symmetric MPS, one can follow the protocol of Ref.~\cite{rupprecht2026faster}, which reorders rows and columns into block-diagonal form and synthesises the blocks in parallel, at a cost dominated by the size of the largest block (rather than the sum of these sizes, i.e. the bond dimension $\chi$). Therefore, as a corollary to Lemma~\ref{lemma:mapping}, preparation at this cost follows trivially:
\begin{corollary}[Preparation of the step-vector MPS]\label{corollary:loading}
Place the spin-adapted MPS in right-canonical gauge. Then:

\textbf{(i)} The matrix of site $i$ in Equation~\eqref{eq:site-unitary} is block-sparse, with one block for each sector $\Gamma$ on bond $i-1$, with $n^{(i-1)}_{\Gamma}$ orthonormal columns, one for each multiplet of the sector, and
\begin{equation}\label{eq:block-height}
    h^{(i)}_{\Gamma} = \sum_{\delta=0}^{3} n^{(i)}_{\Gamma-f(\delta)}
\end{equation}
rows, where $n^{(i)}_{\Gamma'}=0$ for a sector $\Gamma'$ that does not occur on bond $i$. The block-sparse preparation of Ref.~\cite{rupprecht2026faster} therefore applies without modification and prepares the state~\eqref{eq:step-state}, at the Toffoli count of that reference for blocks of these sizes.

\textbf{(ii)} Let the target be a singlet, and let its spin-adapted MPS be minimal, that is, no spin-adapted MPS of the target has fewer multiplets of any sector on any bond. Then the target $\ket{\Psi_{00}}$, written as a right-canonical $\U1\times\U1$-symmetric MPS whose bond dimension on every bond equals the Schmidt rank there, has the structure of (i) with the sectors $\Gamma$ replaced by the charges $(N',m)$ and the numbers of multiplets $n^{(i)}_\Gamma$ replaced by the numbers of states
\begin{equation}\label{eq:state-count}
    c^{(i)}_{(N',m)} = \sum_{S'\geq\lvert m\rvert} n^{(i)}_{(N',S')},
\end{equation}
where the sum runs over the spins with $S'-m$ an integer: one block for each charge of bond $i-1$, with $c^{(i-1)}_{(N',m)}$ orthonormal columns and $\sum_{\sigma} c^{(i)}_{(N'-n_\sigma,\,m-m_\sigma)}$ rows, where $(n_\sigma,m_\sigma)$ is the charge of the occupation state $\sigma$. Counted over all blocks, a multiplet of spin $S'$ accounts for one column in (i) and for $2S'+1$ columns here, one in each of $2S'+1$ blocks, and a block here with $\lvert m\rvert\leq\tfrac12$ holds the multiplets of every spin of its particle number, where a block of (i) holds those of one spin.
\end{corollary}
\begin{proof}
\textbf{(i)} Each row $(\delta,\Gamma_i,\alpha_i)$ of the matrix belongs to exactly one block, that of the sector $\Gamma_{i-1}=\Gamma_i+f(\delta)$. The block of $\Gamma$ therefore stacks, for each of the at most four steps, the block $(R_i^{q_\delta})_{\Gamma,\Gamma-f(\delta)}$ of the reduced tensor, and the orthonormality of its columns is Equation~\eqref{eq:su2-isometry}.
\textbf{(ii)} Fix a bond $i$. The boundary bonds each carry a single spin-zero multiplet, so Equation~\eqref{eq:state-count} holds for them directly. We expand the target across the cut at bond $i$, read off the reduced density matrix of the right block, count its states of each charge, and place these counts in the blocks of the site matrix.

\emph{The target across the cut.} In the right-canonical gauge, by Proposition~\ref{prop:su2-mps}, the right-block states $\ket{\Gamma\alpha;M}$ on bond $i$ are orthonormal, and for each $(\Gamma,\alpha)$ they are a standard basis of a multiplet of particle number $N_\Gamma$ and spin $S_\Gamma$ of the orbitals $i+1,\dots,d$. Complete them to an orthonormal multiplet basis of these orbitals and expand the target in it,
\begin{equation*}
    \ket{\Psi_{00}}=\sum_{\Gamma,\alpha,M}\ket{X_{\Gamma\alpha M}}\otimes\ket{\Gamma\alpha;M},
\end{equation*}
with vectors $\ket{X_{\Gamma\alpha M}}$ on the orbitals $1,\dots,i$; the terms of the completion vanish, since the target is the contraction of the MPS through bond $i$. The orbitals $1,\dots,i$ and $i+1,\dots,d$ are subsystems in the sense of Proposition~\ref{prop:wigner-eckart-theorem} (see the discussion following Equation~\eqref{eq:local-multiplets}). Apply its part (i) with total spin $0$, an orthonormal multiplet basis $\ket{a,S_1,M_1}$ of the orbitals $1,\dots,i$ as the first subsystem and the right block as the second. The triangular condition with total spin $0$ forces $S_1=S_\Gamma$, and projection conservation gives $M_1=-M$, so
\begin{equation*}
    \ket{X_{\Gamma\alpha M}} = \cg{S_\Gamma}{-M}{S_\Gamma}{M}{0}{0}\sum_{a}\tilde C^{\Gamma}_{a\alpha}\ket{a,S_\Gamma,-M},
\end{equation*}
where the matrix $\tilde C^{\Gamma}$ does not depend on $M$ and has $n^{(i)}_\Gamma$ columns. The sum runs over the multiplets $a$ of spin $S_\Gamma$ and particle number $N-N_\Gamma$, since the target has particle number $N$ and $\ket{\Gamma\alpha;M}$ has $N_\Gamma$.

\emph{The reduced density matrix of the right block.} In the two orthonormal bases $\ket{a,S_1,M_1}$ and $\ket{\Gamma\alpha;M}$ the coefficients of the target are block-diagonal in $(\Gamma,M)$, with the block $\cg{S_\Gamma}{-M}{S_\Gamma}{M}{0}{0}\tilde C^\Gamma$. Since $\lvert\cg{S}{-M}{S}{M}{0}{0}\rvert^2=(2S+1)^{-1}$, the Schmidt values on bond $i$ are the singular values of $\tilde C^\Gamma$ divided by $(2S_\Gamma+1)^{1/2}$, each repeated for the $2S_\Gamma+1$ projections $M$. With the squared Clebsch-Gordan coefficient equal to $(2S_\Gamma+1)^{-1}$ and the states $\ket{a,S_1,M_1}$ orthonormal, the reduced density matrix $\rho$ of the orbitals $i+1,\dots,d$ has, in the basis $\ket{\Gamma\alpha;M}$, the entries
\begin{equation}
    \begin{aligned}
    \rho_{(\Gamma\alpha M),(\Gamma'\alpha'M')}&=\braket{X_{\Gamma'\alpha'M'}}{X_{\Gamma\alpha M}}\\ &=\frac{\delta_{\Gamma\Gamma'}\delta_{MM'}}{2S_\Gamma+1}\big(\tilde C^{\Gamma\dagger}\tilde C^{\Gamma}\big)_{\alpha'\alpha}.
\end{aligned}
\end{equation}

It is block-diagonal, with one block for each sector $\Gamma$ and projection $M$, and vanishes on the states of the completion. Each block has the rank of $\tilde C^\Gamma$, which is $n^{(i)}_\Gamma$ since the spin-adapted MPS is minimal. Suppose instead that $\tilde C^\Gamma$ had rank $r<n^{(i)}_\Gamma$. There is then a unitary $V$ such that the last $n^{(i)}_\Gamma-r$ columns of $\tilde C^\Gamma V$ vanish. Replacing $\ket{\Gamma\beta;M}$ by $\sum_\alpha V^*_{\alpha\beta}\ket{\Gamma\alpha;M}$, with the same $V$ for every $M$, changes the coefficient matrix to $\tilde C^\Gamma V$ and leaves the spin multiplets intact. This is a gauge transformation on bond $i$, absorbed into the two adjacent reduced tensors; no other bond changes. The multiplets whose vectors $\ket{X}$ vanish can then be deleted from the bond without changing the target. This contradicts minimality.

\emph{The states of each charge.} The states $\ket{\Gamma\alpha;M}$ of charge $(N',m)$ are those with $N_\Gamma=N'$ and $M=m$. A sector $\Gamma=(N',S')$ contains such states exactly when $S'\geq\lvert m\rvert$ and $S'-m$ is an integer, and then $n^{(i)}_\Gamma$ of them, so the restriction of $\rho$ to the charge $(N',m)$ has rank $c^{(i)}_{(N',m)}$. In the $\U1\times\U1$-symmetric MPS of the statement, the right-block states of bond $i$ are orthonormal by right-canonicality and have definite charges (Definition~\ref{def:symmetric-mps}). Their span contains the support of $\rho$, since the target is the contraction of the MPS through bond $i$. Its dimension is the bond dimension, and hence the Schmidt rank, which is the dimension of the support. The span is therefore the support. The charge operators of the orbitals $i+1,\dots,d$ commute with $\rho$, so the rank of $\rho$ on the charge $(N',m)$ is the number of basis states of that charge. This is Equation~\eqref{eq:state-count}.

\emph{The blocks of the site matrix.} Since $i$ was arbitrary, the same holds on bond $i-1$. The block-sparse preparation of Ref.~\cite{rupprecht2026faster} groups the columns of the site matrix by the charge of bond $i-1$, so the block of the charge $(N',m)$ has $c^{(i-1)}_{(N',m)}$ columns, orthonormal because the $\U1\times\U1$ MPS is right-canonical. By the selection rule~\eqref{eq:charge-cons} with bonds labelled by the orbitals to their right, a row $(\sigma,\alpha_i)$ belongs to the block whose charge is that of $\alpha_i$ plus $(n_\sigma,m_\sigma)$, so the block of $(N',m)$ has $\sum_\sigma c^{(i)}_{(N'-n_\sigma,\,m-m_\sigma)}$ rows, where $c^{(i)}=0$ for a charge absent from bond $i$. Ref.~\cite{rupprecht2026faster} labels each bond by the charge $(N_L,m_L)$ of the orbitals to its left, which is $(N-N_R,-m_R)$ for our right charge $(N_R,m_R)$, so the two labellings differ by a one-to-one relabelling of the blocks of the site matrix.

\emph{Counting multiplets.} Summing Equation~\eqref{eq:state-count} over $m$ counts each multiplet of spin $S'$ once for each of its $2S'+1$ projections, in $2S'+1$ different charges. Finally, each $\ket{\Gamma\alpha;M}$ is a non-zero state of $N_\Gamma$ electrons, and each electron contributes $\pm\tfrac12$ to the spin projection, so $M-\tfrac12N_\Gamma$, and with it $S_\Gamma-\tfrac12N_\Gamma$, is an integer. Every charge $(N',m)$ that occurs has $m-\tfrac12N'$ an integer. A charge with $\lvert m\rvert\leq\tfrac12$ has $m=0$ when $N'$ is even and $m=\pm\tfrac12$ when $N'$ is odd, and in either case the sum~\eqref{eq:state-count} runs over all the spins of the sectors of particle number $N'$, since these are integers for even $N'$ and half-integers of at least $\tfrac12$ for odd $N'$.
\end{proof}

We briefly remark that abelian point-group symmetries of the Hamiltonian can be incorporated in the same manner, as one more entry of the sector label. Every irreducible representation of an abelian point group is one-dimensional, and the orbitals can be chosen such that orbital $i$ transforms as one of them, $g_i$. A state of the block to the right of bond $i$ then transforms as a single irreducible representation $I_i$, and attaching orbital $i$ with $\Delta N$ electrons gives $I_{i-1}=I_i\,g_i^{\Delta N}$. Since a step fixes the occupation of its orbital (Table~\ref{tab:steps}), it also fixes this change. The sector can thus be enlarged to $\Gamma=(N,S,I)$ (the standard labelling used by e.g. \texttt{block2} with point-group symmetries enabled~\cite{zhai2023block2}), and Lemma~\ref{lemma:mapping} and Corollary~\ref{corollary:loading} hold for it with a site-dependent increment in place of $f(\delta)$, each block of the site matrix being subdivided into smaller ones. The inverse Clebsch-Gordan transforms of Section~\ref{sec:cg-step} are unchanged, since the point group acts on the spatial orbitals alone and commutes with the total spin.

\subsection{From step vectors to occupations}\label{sec:cg-step}
From Corollary~\ref{corollary:loading}, we now know that we can prepare the state $\ket{\Phi}$ of the step-vector MPS efficiently. Observing Equation~\ref{eq:csf-expansion}, what remains to prepare the desired state $\ket{\Psi_{SM}}$ is implementing the linear map $U_M$ with $U_M\ket{\delta_1}\cdots\ket{\delta_d}=\ket{N,S,M;\bm\delta}$, defined on the span of the legal step vectors. Since the CSFs of fixed $M$ are orthonormal, $U_M$ is an isometry from this span onto the states with quantum numbers $(N,S,M)$. Up to a reversal of orbital coupling order, it is the inverse of the quantum Paldus transform of Ref.~\cite{burkat2025quantum}; the construction of the transformation is based on their approach.

First, we define the tensor $K$ that enacts this transform site-wise by coupling the spin $S''$ of the right block of orbitals to an orbital in the multiplet $q_\delta$, giving the spin $S'=S''+\Delta S(\delta)$:
\begin{equation}\label{eq:step-coupling}
    \big(K^{\delta,(qm)}\big)_{(S'M'),(S''M'')}
    = \begin{cases}
        \cg{S''}{M''}{s}{m}{S'}{M'} & \text{if } q=q_\delta \text{ and} \\
        & \quad S'=S''+\Delta S(\delta), \\
        0 & \text{otherwise,}
    \end{cases}
\end{equation}
where $q=(n,s)$ and $\Delta S(\delta)$ is the change of spin of the step $\delta$ (Table~\ref{tab:steps}). The tensor is site- and state-independent, and with this we may rewrite Equation~\ref{eq:csf-expansion} as:
\begin{equation}\label{eq:diag-two-layer}
\begin{gathered}
\psi_{SM}^{\vec{q},\vec{m}} \;=\;
\begin{tikzpicture}[baseline={([yshift=-.5ex]0,0.45)},
  cgleg/.style={line width=1},
  cglabel/.style={font=\scriptsize,inner sep=2pt},
  cgdots/.style={font=\scriptsize,fill=white,inner sep=1.5pt},
  cgtip/.pic={\draw[line width=1,-{Stealth[length=7pt,width=4pt]}] (-3.5pt,0) -- (3.5pt,0);},
  cgmid/.style={pos=0.5,sloped,allow upside down}]
% tikz-network edges in the colour of the surrounding text, like the plain legs
\colorlet{cgline}{.}
% upper row: step-vector tensors. Lower row: the four-leg tensor K of eq:step-coupling, one below
% each step-vector tensor. K is not the three-leg Clebsch-Gordan node, so it is drawn as a square.
\Vertex[x=0,y=0.9,label={$A_1$},position=center,size=0.5,color=red!30]{A1}
\Vertex[x=1.6,y=0.9,label={$A_2$},position=center,size=0.5,color=red!30]{A2}
\Vertex[x=3.9,y=0.9,label={$A_d$},position=center,size=0.5,color=red!30]{Ad}
\Vertex[x=0,y=0,label={$K$},position=center,size=0.42,color=gray!30,shape=rectangle]{C1}
\Vertex[x=1.6,y=0,label={$K$},position=center,size=0.42,color=gray!30,shape=rectangle]{C2}
\Vertex[x=3.9,y=0,label={$K$},position=center,size=0.42,color=gray!30,shape=rectangle]{Cd}
% multiplet lines of the step-vector MPS
\Edge[lw=1,color=cgline](A1)(A2)
\Edge[lw=1,color=cgline](A2)(Ad)
\node[cglabel,above] at (0.8,0.9) {$\Gamma_1\alpha_1$};
\node[cgdots] at (2.75,0.9) {$\cdots$};
\draw[cgleg] (A1) -- (-0.65,0.9) node[cglabel,left]{$N\,S$};
\draw[cgleg] (Ad) -- (4.55,0.9) node[cglabel,right]{$0\,0$};
% step lines: the only lines that join the two rows
\Edge[lw=1,color=cgline](A1)(C1)
\Edge[lw=1,color=cgline](A2)(C2)
\Edge[lw=1,color=cgline](Ad)(Cd)
\node[cglabel,right] at (0,0.42) {$\delta_1$};
\node[cglabel,right] at (1.6,0.42) {$\delta_2$};
\node[cglabel,right] at (3.9,0.42) {$\delta_d$};
% spin lines, from the vacuum on the right to the total spin on the left
\Edge[lw=1,color=cgline](C2)(C1)
\Edge[lw=1,color=cgline](Cd)(C2)
\path (C2) -- pic[cgmid]{cgtip} (C1);
\path (Cd) -- pic[cgmid,pos=0.22]{cgtip} (C2);
\node[cglabel,below,inner sep=4pt] at (0.8,0) {$S_1M_1$};
\node[cgdots] at (2.75,0) {$\cdots$};
\draw[cgleg] (4.55,0) node[cglabel,right]{$0\,0$} -- pic[cgmid]{cgtip} (Cd);
\draw[cgleg] (C1) -- pic[cgmid]{cgtip} (-0.65,0) node[cglabel,left]{$S\,M$};
% orbitals: second input of each node, from below
\draw[cgleg] (0,-0.62) node[cglabel,below]{$q_1m_1$} -- pic[cgmid]{cgtip} (C1);
\draw[cgleg] (1.6,-0.62) node[cglabel,below]{$q_2m_2$} -- pic[cgmid]{cgtip} (C2);
\draw[cgleg] (3.9,-0.62) node[cglabel,below]{$q_dm_d$} -- pic[cgmid]{cgtip} (Cd);
\end{tikzpicture}
\end{gathered}.
\end{equation}
We apply $U_M$ one site at a time, interleaved with the site isometries of Section~\ref{sec:loading}: the $K$ of site $i$ acts directly after the site isometry $A_i$ of orbital $i$, while the bond register holds the sector $\Gamma_i$.

To implement this on a quantum computer, we need an efficient unitary that acts on the registers of the circuit. At site $i$, $K$ as written does not define a map on these registers: its entries depend on the projections $M_{i-1}$ and $M_i$ carried by the bonds on either side of orbital $i$, and these are not held by any register yet (the bond register holds only the sector $\Gamma_i=(N_i,S_i)$, and a bond index stands for all $2S_i+1$ projections of its multiplet by construction). We therefore add an $M$ register to hold the projection at the current bond. It is initialised on the leftmost bond in the state $\ket{M}$, the projection of the target, since bond $0$ carries the spin of the whole chain. Site $i$ subtracts the projection $m_i$ of its orbital, so that the register contains $M_i=M_{i-1}-m_i$ after site $i$ and $M_d=0$ at the end. We do not require such a construction for total spin since the angle depends only on $S_i$, which is read from the bond register, and on $M_{i-1}$, and the step itself dictates whether the spin increments or decrements.

The initialisation of the $M$ register is the reason for our right-to-left direction of spin fusion: the starting bond is the only place where the projection $M$ of the whole state can be set, and starting from the vacuum end would leave the system qubits entangled with the $M$ register at the end. Spin fusion most conventionally flows left to right~\cite{paldus1974group,shepard2005general,weichselbaum2012non,keller2016spin}, but for MPS state preparation, isometries are usually chosen to dilate left-to-right (right-canonical gauge)~\cite{schoen2005sequential,fomichev2024initial,berry2025rapid,rupprecht2026faster} --- however these must flow in opposite directions due to this need to set projection at the starting bond, forcing us to choose a reversed convention for one of these. As this is a paper about state preparation, we elected to use left-to-right state preparation and right-to-left spin fusion; the opposite order can be achieved by a straightforward reversal of our conventions. As noted in Appendix~\ref{app:conventions}, a \texttt{block2} calculation ending on a right-to-left sweep delivers its tensors in this form for free; care must be taken when interfacing with spin-adapted DMRG codes to ensure this is so, as conventions differ greatly from library to library.

Now for the construction of the map itself, per Ref.~\cite{burkat2025quantum}. Recall that the two qubits of orbital $i$ track the occupation of its modes $(i,\uparrow)$ and $(i,\downarrow)$ in a Jordan-Wigner mapping, with the occupation states $\ket{\emptyset}=\ket{00}$, $\ket{\uparrow}=\ket{10}$, $\ket{\downarrow}=\ket{01}$ and $\ket{\uparrow\downarrow}=\ket{11}$. Prior to the transform, we write the step on the same two qubits, as $0\mapsto\ket{00}$, $1\mapsto\ket{10}$, $2\mapsto\ket{01}$ and $3\mapsto\ket{11}$. This map is trivial for empty ($\delta=0$) or doubly-occupied ($\delta=3$) orbitals, both of which have $s_i=0$, and neither the total spin nor its projection change across the bond, so the Clebsch-Gordan coefficient is $1$ and the transform is the identity:
\begin{equation}\label{eq:cg-trivial}
\begin{aligned}
    \ket{\delta{=}0}\ket{M_{i-1}} &\mapsto \ket{\emptyset}\ket{M_{i-1}},\\
    \ket{\delta{=}3}\ket{M_{i-1}} &\mapsto \ket{\uparrow\downarrow}\ket{M_{i-1}},
\end{aligned}
\end{equation}
where the second ket denotes the $M$ register. For singly-occupied orbitals ($\delta\in\{1,2\}$) the electron raises or lowers the spin, $S_{i-1}=S_i\pm\tfrac12$. We are only adding one spin-$\tfrac12$ at a time, so the full form of the Racah formula is not needed --- the Clebsch-Gordan coefficients are simple geometric factors, the $M$ register drops by the orbital's projection, and the remaining cases of the transform thus are given by
\begin{equation}\label{eq:cg-rotation}
\begin{aligned}
    \ket{\delta{=}1}\ket{M_{i-1}} &\mapsto \cos\theta\,\ket{\uparrow}\ket{M_{i-1}-\tfrac12} \\
    &\quad+ \sin\theta\,\ket{\downarrow}\ket{M_{i-1}+\tfrac12},\\
    \ket{\delta{=}2}\ket{M_{i-1}} &\mapsto -\sin\theta\,\ket{\uparrow}\ket{M_{i-1}-\tfrac12} \\
    &\quad+ \cos\theta\,\ket{\downarrow}\ket{M_{i-1}+\tfrac12},
\end{aligned}
\end{equation}
where the geometric factors for the Clebsch-Gordan coefficients are given by
\begin{equation}\label{eq:cg-angle}
    \cos\theta = \sqrt{\frac{S_i+M_{i-1}+\tfrac12}{2S_i+1}},\quad
    \sin\theta = \sqrt{\frac{S_i-M_{i-1}+\tfrac12}{2S_i+1}}.
\end{equation}
We denote the linear map enacted in Equations~\ref{eq:cg-trivial} and~\ref{eq:cg-rotation} by $\hat K_{S_i}$, which should be thought of as the unitary completion of the tensor $K$ such that it can be implemented on a quantum computer. It is the inverse of the Clebsch-Gordan transform of Ref.~\cite{burkat2025quantum}, with $S_i$ read from the bond register in place of a spin register:
\begin{lemma}[Inverse Clebsch-Gordan transform~\cite{burkat2025quantum}]\label{lemma:cg-transform}
Choose a spin $S_i\geq0$, and restrict $\hat K_{S_i}$ to act on the span of states $\ket{\delta}\ket{M_{i-1}}$ for which the spin $S_{i-1}=S_i+\Delta S(\delta)$ is non-negative and $M_{i-1}\in\{-S_{i-1},\dots,S_{i-1}\}$. Then:

\textbf{(i)} The amplitudes of $\hat K_{S_i}$ are the entries of the tensor $K$ of Equation~\eqref{eq:step-coupling}:
\begin{equation}\label{eq:cg-transform-entries}
    \hat K_{S_i}\ket{\delta}\ket{M_{i-1}} = \sum_{m}\cg{S_i}{M_{i-1}-m}{s}{m}{S_{i-1}}{M_{i-1}}\,\ket{q_\delta,m}\ket{M_{i-1}-m},
\end{equation}
where $s$ is the spin of $q_\delta$ and $\ket{q_\delta,m}$ is the local state of Equation~\eqref{eq:local-multiplets} on the two qubits of the orbital.

\textbf{(ii)} $\hat K_{S_i}$ is an isometry.
\end{lemma}

\begin{proof}
\textbf{(i)} $\hat K_{S_i}$ is the inverse of the Clebsch-Gordan transform given in Section III of Ref.~\cite{burkat2025quantum} for an orbital coupled to a block of spin $S_i$, without its spin and particle-number registers; its coefficients (Equation 12, Ref~\cite{burkat2025quantum}) are those of Definition~\ref{def:cg}, with the block's spin first and no change of sign, which yields the desired result \textbf{(ii)} This holds trivially because the update of the $M$ register and the rotations of that transform are unitary.
\end{proof}

\subsection{The full preparation circuit}\label{sec:preparation}
We now have the two tools needed for the full circuit construction: Corollary~\ref{corollary:loading} loads the reduced tensors, in the form of the step-vector MPS, and Lemma~\ref{lemma:cg-transform} returns the step of a site to the occupation basis. In this section we assemble them into the full preparation circuit and prove that it prepares the corresponding quantum state of the spin-adapted MPS.

The idea is as follows. For each $M$, the target $\ket{\Psi_{SM}}$ is an ordinary MPS (Definition~\ref{def:mps}) with isometric site tensors $T_1,\dots,T_d$ (Equation~\eqref{eq:site-factorisation}), and the circuit sequentially applies these to the system register, with the bond index $(\Gamma_i\alpha_iM_i)$ held by two registers: $(\Gamma_i\alpha_i)$ by the bond register and $M_i$ by the separate $M$ register. Each $T_i$ factorises into a reduced tensor $R_i$ times one Clebsch-Gordan coefficient $\mathcal{C}$ (the Wigner-Eckart theorem), and the two factors are applied by separate operations: the site isometry of the step-vector MPS supplies the reduced tensor's amplitudes, whilst the inverse Clebsch-Gordan transform that follows it supplies the purely geometric Clebsch-Gordan coefficient.
The circuit runs from left to right, splitting off one orbital at a time, until the system qubits are in the state $\ket{\Psi_{SM}}$ and only the vacuum is left on bond $d$, unentangled from the prepared state.

\begin{figure}[t]
\centering
\begin{tikzpicture}[x=0.82cm,y=0.8cm,line width=0.6pt]
\draw (0,2) -- (6.2,2);
\draw (0,1) -- (6.2,1);
\draw (0,0) -- (6.2,0);
\node[left] at (0,2) {\scriptsize$\ket{\Gamma_{i-1}\alpha_{i-1}}_{\mathrm{bond}}$};
\node[left] at (0,1) {\scriptsize$\ket{00}$};
\node[left] at (0,0) {\scriptsize$\ket{M_{i-1}}$};
\node[right] at (6.2,2) {\scriptsize$\ket{\Gamma_{i}\alpha_{i}}_{\mathrm{bond}}$};
\node[right] at (6.2,1) {\scriptsize orbital $i$};
\node[right] at (6.2,0) {\scriptsize$\ket{M_{i}}$};
\draw[fill=white] (0.6,0.55) rectangle (1.8,2.45);
\node at (1.2,1.5) {\scriptsize$A_i$};
\node[above] at (2.45,1) {\scriptsize$\ket{\delta_i}$};
\draw (3.6,2) -- (3.6,0);
\draw[fill=white] (3.1,0.6) rectangle (4.1,1.4);
\node at (3.6,1) {\scriptsize$\theta$};
\draw[fill=white] (3.44,1.84) rectangle (3.76,2.16);
\draw[fill=white] (3.44,-0.16) rectangle (3.76,0.16);
\node[above] at (3.6,2.16) {\tiny$S_i$};
\node[below] at (3.6,-0.16) {\tiny$M_{i-1}$};
\draw (5.2,1) -- (5.2,0);
\fill (5.2,1) circle (2pt);
\draw[fill=white] (4.7,-0.38) rectangle (5.7,0.38);
\node at (5.2,0) {\scriptsize$-m_i$};
\node at (1.2,2.95) {\scriptsize(a)};
\draw (3.1,2.7) -- (3.1,2.8) -- (5.7,2.8) -- (5.7,2.7);
\node at (4.4,3.05) {\scriptsize(b)};
\end{tikzpicture}
\caption{\label{fig:site-circuit}\textbf{One site of the preparation circuit.} The wires depict the bond register (top), the two qubits representing orbital $i$ (middle) and the $M$ register (bottom). (a) The site isometry $A_i$ of the step-vector MPS (Equation~\eqref{eq:site-unitary}) advances the bond register from bond $i-1$ to bond $i$ and leaves the two qubits in a superposition of the encoded steps $\ket{\delta_i}$. (b) The inverse Clebsch-Gordan transform rotates the encoded steps $1$ and $2$ by the angle $\theta$ of Equation~\eqref{eq:cg-angle}, multiplexed over the spin $S_i$ read from the bond register and the value $M_{i-1}$ of the $M$ register, and then subtracts the projection $m_i$ of the orbital from the $M$ register.}
\end{figure}
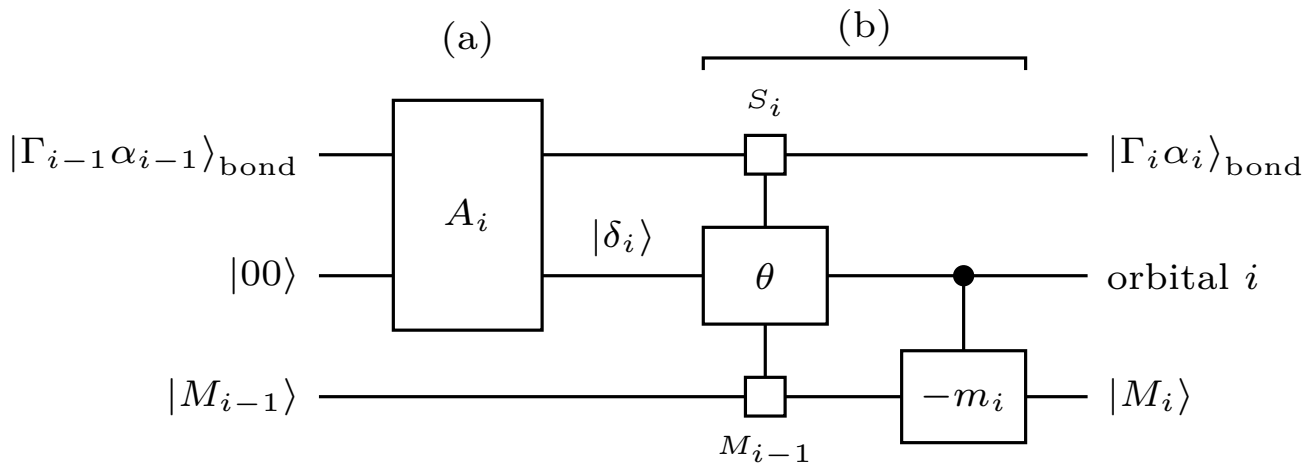

In Figure~\ref{fig:site-circuit}, we illustrate one site of the resulting preparation circuit.
The circuit acts on the bond register, the $M$ register, and the $2d$ qubits of the orbitals. Initially, the bond register holds the single trivial multiplet of the target sector $\Gamma_0=(N,S)$, the $M$ register holds the desired projection $M$, and the qubits are in $\ket{0}$. For $i=1,2,\dots,d$ the circuit then applies, orbital-by-orbital:
\begin{enumerate}
    \item the site isometry~\eqref{eq:site-unitary} of the step-vector MPS on the bond register and the two qubits of orbital $i$, with $A_i^\delta$ as the site tensor and the step encoded as in Section~\ref{sec:cg-step};
    \item the inverse Clebsch-Gordan transform $\hat K_{S_i}$ on the two qubits of orbital $i$ and the $M$ register, controlled on the spin $S_i$ read from the bond register.
\end{enumerate}
We can now state the main result of this work: the preparation circuit prepares the spin-adapted MPS for any projection $M$, at the cost of loading its step-vector MPS (typically lower than that of Ref.~\cite{rupprecht2026faster}, since it is determined by $\chi$ rather than $\chi^{\ast}$), plus a small overhead at each site that does not depend on the entries of the tensors. This overhead depends on the bond dimension only through the read-out of the spin from the bond register, which a spin register of $\lceil\log_2(d+1)\rceil$ qubits removes (Section~\ref{sec:cost}).
\begin{theorem}[Preparation of a spin-adapted MPS]\label{thm:preparation}
Let $R_1,\dots,R_d$ be the reduced tensors of a right-canonical spin-adapted MPS (Definition~\ref{def:su2-mps}) representing the multiplet $\ket{\Psi_{SM}}$, $M=-S,\dots,S$. Then:

\textbf{(i)} For every $M$, the preparation circuit ends in the state $\ket{\Gamma_d}_{\mathrm{bond}}\ket{0}_M\otimes\ket{\Psi_{SM}}$, in which the qubits hold $\ket{\Psi_{SM}}$ in the Jordan-Wigner encoding of Section~\ref{sec:problem-setup} and the ancillary registers are in states that do not depend on $M$.

\textbf{(ii)} Every operation of the circuit is an isometry, and so extends to a unitary.

\textbf{(iii)} The site isometries are those of Corollary~\ref{corollary:loading}, with the same blocks and the same cost. At each site the inverse Clebsch-Gordan transform adds one rotation multiplexed over the allowed values of $(S_i,M_{i-1})$, one controlled update of the $M$ register, and the read-out of $S_i$ from the bond register with its uncomputation. None of these depends on the entries of the tensors; the cost of the read-out depends on how the bond register is laid out and on $\chi_i$ (Section~\ref{sec:cost}).
\end{theorem}
The proof uses the fact that the qubits hold products of orbital and block states without Jordan-Wigner strings. It expands the target at every cut of the chain and shows by induction that after site $i$ the registers hold the expansion at bond $i$, with each state of the orbitals not yet prepared replaced by its labels. Parts (ii) and (iii) then follow from Lemmas~\ref{lemma:mapping} and~\ref{lemma:cg-transform} and Corollary~\ref{corollary:loading}.
\begin{proof}
Throughout, for sectors $\Gamma_{i-1}$ and $\Gamma_i$ of neighbouring bonds, $\delta(\Gamma_{i-1},\Gamma_i)$ is the step with $\Gamma_{i-1}=\Gamma_i+f(\delta)$, and $q=(n,s)$ is its local multiplet. When it exists, the step is unique, since the four increments $f(\delta)$ are distinct. When no step joins the two sectors, every block $(R_i^{q})_{\Gamma_{i-1}\Gamma_i}$ vanishes by the selection rule~\eqref{eq:su2-selection}, and the pair vanishes from the sums below. Two preliminaries come first:

\textbf{Qubit form of the products.}
By Section~\ref{sec:fermions}, the product $\ket{q,m}\otimes\ket{y}$ of a local state of orbital $i$ and a state $\ket{y}$ of the orbitals $i+1,\dots,d$ is encoded as the tensor product of their qubit states, with no Jordan-Wigner strings. The same holds for every product below, since in each the operators are ordered by increasing orbital index.

\textbf{The target expanded at a cut.}
We first write the target in the form the circuit holds after site $i$: the orbitals $1,\dots,i$ in vectors $\ket{\omega^{(i)}}$, entangled with the multiplets of the orbitals $i+1,\dots,d$, which the circuit keeps in the bond and $M$ registers until it reaches them (Part (i)).
For $i=0,\dots,d$, we define vectors $\ket{\omega^{(i)}_{\Gamma_i\alpha_i M_i}}$ on the left orbitals $1,\dots,i$ recursively. For $i=0$ there are no orbitals, the bond has its single multiplet, and we set $\omega^{(0)}_{\Gamma_0 1 M_0}=\delta_{M_0 M}$. For $i\geq1$,
\begin{equation}\label{eq:omega-recursion}
\begin{aligned}
    \ket{\omega^{(i)}_{\Gamma_i\alpha_i M_i}} &= \sum_{\Gamma_{i-1},\alpha_{i-1}} \sum_{m} (R_i^{q})_{(\Gamma_{i-1}\alpha_{i-1}),(\Gamma_i\alpha_i)} \cg{S_i}{M_i}{s}{m}{S_{i-1}}{M_i+m} \\
    &\qquad\times\ket{\omega^{(i-1)}_{\Gamma_{i-1}\alpha_{i-1},M_i+m}}\otimes\ket{q,m},
\end{aligned}
\end{equation}
with $q$ the local multiplet of the step $\delta(\Gamma_{i-1},\Gamma_i)$. Then, for every $i$,
\begin{equation}\label{eq:cut-expansion}
    \ket{\Psi_{SM}} = \sum_{\Gamma_i,\alpha_i,M_i} \ket{\omega^{(i)}_{\Gamma_i\alpha_i M_i}}\otimes\ket{\Gamma_i\alpha_i;M_i}.
\end{equation}
For $i=0$ this is $\ket{\Psi_{SM}}=\ket{\Gamma_0,1;M}$, the end of the recursion~\eqref{eq:su2-states}. If it holds for $i-1$, substitute Equations~\eqref{eq:su2-states} and~\eqref{eq:site-coupling} for each right-block state $\ket{\Gamma_{i-1}\alpha_{i-1};M_{i-1}}$. By the selection rule~\eqref{eq:su2-selection}, for given $\Gamma_{i-1}$ and $\Gamma_i$ the only local multiplet that contributes is that of the step $\delta(\Gamma_{i-1},\Gamma_i)$. Renaming $M_{i-1}-m$ as $M_i$, the coefficient of $\ket{\Gamma_i\alpha_i;M_i}$ is the right-hand side of Equation~\eqref{eq:omega-recursion}, which gives the statement for $i$. At $i=d$ the right block is empty, the only term is the vacuum with $M_d=0$, and Equation~\eqref{eq:cut-expansion} reads $\ket{\Psi_{SM}}=\ket{\omega^{(d)}_{\Gamma_d 1 0}}$.

\textbf{Part (i): the state after each site.}
After site $i$ the registers hold the expansion~\eqref{eq:cut-expansion} of the target at bond $i$, with each right-block state $\ket{\Gamma_i\alpha_i;M_i}$, a state of orbitals the circuit has not yet reached, replaced by its labels: $(\Gamma_i\alpha_i)$ in the bond register and $M_i$ in the $M$ register. That is, after the operations of sites $1,\dots,i$ the registers hold
\begin{equation}\label{eq:invariant}
    \sum_{\Gamma_i,\alpha_i,M_i} \ket{\Gamma_i\alpha_i}_{\mathrm{bond}}\,\ket{M_i}_M\,\ket{\omega^{(i)}_{\Gamma_i\alpha_i M_i}}_{1\dots i}\,\ket{0}_{i+1\dots d}.
\end{equation}
We prove this by induction on $i$. For $i=0$ it is the initial state, since $\omega^{(0)}_{\Gamma_0 1 M_0}=\delta_{M_0M}$. Assume it for $i-1$, and follow one term, with labels $(\Gamma_{i-1}\alpha_{i-1})$ and $M_{i-1}$, through the two operations of site $i$.

\emph{The site isometry} writes the step, with the entries of the reduced tensor as amplitudes. By Equations~\eqref{eq:site-unitary} and~\eqref{eq:step-tensor} it maps $\ket{\Gamma_{i-1}\alpha_{i-1}}_{\mathrm{bond}}\ket{00}$ to
\begin{equation*}
    \sum_{\Gamma_i,\alpha_i}(R_i^{q})_{(\Gamma_{i-1}\alpha_{i-1}),(\Gamma_i\alpha_i)}\,\ket{\Gamma_i\alpha_i}_{\mathrm{bond}}\ket{\delta},
\end{equation*}
with $\delta=\delta(\Gamma_{i-1},\Gamma_i)$ and $q$ its local multiplet.

\emph{The inverse Clebsch-Gordan transform} turns the step into the state of the orbital and subtracts the projection $m$ of that state from the $M$ register. It applies $\hat K_{S_i}$, with $S_i$ read from the bond register, and by Lemma~\ref{lemma:cg-transform}(i) this maps $\ket{\delta}\ket{M_{i-1}}_M$ to
\begin{equation*}
    \sum_m \cg{S_i}{M_{i-1}-m}{s}{m}{S_{i-1}}{M_{i-1}}\,\ket{q,m}\ket{M_{i-1}-m}_M .
\end{equation*}
The spin $S_i+\Delta S(\delta)$ of the lemma is the spin $S_{i-1}$ of the sector $\Gamma_{i-1}$, since $\delta=\delta(\Gamma_{i-1},\Gamma_i)$.

The lemma requires $M_{i-1}\in\{-S_{i-1},\dots,S_{i-1}\}$, and every non-zero term meets this. First, on every bond $j$, $\omega^{(j)}_{\Gamma_j\alpha_jM_j}$ is zero when $\lvert M_j\rvert>S_j$: $\omega^{(0)}$ is non-zero only at $M_0=M$, with $\lvert M\rvert\leq S$, and for $j\geq1$ every Clebsch-Gordan coefficient in the recursion~\eqref{eq:omega-recursion} is then zero. Second, $S_{i-1}-M_{i-1}$ is an integer, since $S-M$ is and each site changes the spin and the projection by amounts that differ by an integer.

Summing over the terms and renaming $M_{i-1}-m$ as $M_i$, the coefficient of $\ket{\Gamma_i\alpha_i}_{\mathrm{bond}}\ket{M_i}_M$ is the right-hand side of Equation~\eqref{eq:omega-recursion}, held on the qubits of the orbitals $1,\dots,i$ in the qubit form above. This is Equation~\eqref{eq:invariant} for $i$. At $i=d$ the only sector is the vacuum, with $S_d=0$, so only the term with $M_d=0$ is non-zero: the bond register holds $\ket{\Gamma_d}_{\mathrm{bond}}$, the $M$ register holds $0$, and the qubits hold $\ket{\omega^{(d)}_{\Gamma_d 1 0}}$, which is $\ket{\Psi_{SM}}$ by Equation~\eqref{eq:cut-expansion} at $i=d$.
Since the expansion~\eqref{eq:cut-expansion} of the target at bond $i$ and the state of the circuit after site $i$ obey the same recursion~\eqref{eq:omega-recursion}, the circuit ends in the target.

\textbf{Part (ii): isometries.}
The circuit can be built from unitaries because each of its two operations is an isometry.

\emph{The site isometry} preserves inner products by construction because the step-vector MPS is right-canonical. By Equation~\eqref{eq:site-unitary}, it maps two basis states $\ket{\Gamma_{i-1}\alpha_{i-1}}_{\mathrm{bond}}\ket{00}$ and $\ket{\Gamma'_{i-1}\alpha'_{i-1}}_{\mathrm{bond}}\ket{00}$ to combinations of the orthonormal states $\ket{\Gamma_i\alpha_i}_{\mathrm{bond}}\ket{\delta}$, with the entries of the $A_i^\delta$ as coefficients. The inner product of the two images is therefore the entry of $\sum_\delta A_i^\delta(A_i^\delta)^\dagger$ between $(\Gamma_{i-1}\alpha_{i-1})$ and $(\Gamma'_{i-1}\alpha'_{i-1})$. The spin-adapted MPS is right-canonical, so by Lemma~\ref{lemma:mapping}(iii) this matrix is the identity.

\emph{The inverse Clebsch-Gordan transform} preserves inner products because $\hat K_{S_i}$ does for each spin $S_i$. It leaves the bond register unchanged and applies $\hat K_{S_i}$ for each value of $S_i$ held there. By Lemma~\ref{lemma:cg-transform}(ii), $\hat K_{S_i}$ is an isometry on the domain of that lemma, and by Part (i) the states the circuit reaches lie in that domain. States of the bond register with different $S_i$ are orthogonal, and stay orthogonal because the register is unchanged.

\textbf{Part (iii): cost.}
The cost of the circuit is that of Corollary~\ref{corollary:loading} for the site isometries, plus a cost at each site for the inverse Clebsch-Gordan transform that does not depend on the entries of the tensors.

\emph{The site isometry} is the map~\eqref{eq:site-unitary} for the step-vector MPS, with the step encoded on the two qubits of orbital $i$, so Corollary~\ref{corollary:loading} applies to it.

\emph{The inverse Clebsch-Gordan transform} is, by Equations~\eqref{eq:cg-trivial} and~\eqref{eq:cg-rotation}, one rotation of the encoded steps $1$ and $2$ by the angle~\eqref{eq:cg-angle}, multiplexed over $S_i$ and $M_{i-1}$, followed by one update of the $M$ register controlled on the two qubits of orbital $i$. The angles do not involve the tensors. The control $S_i$ is the spin of the sector of the bond index, a function of the index alone, so it can be computed into an ancilla before the rotation and uncomputed after it. This read-out has a cost that depends on the layout of the bond register and on $\chi_i$, but not on the entries of the tensors.
\end{proof}
We make two remarks. First, besides the states $\ket{\Psi_{SM}}$ of fixed $M$, the circuit can prepare arbitrary superpositions of different $M$ values by initialising the $M$ register accordingly. Part (i) of Theorem~\ref{thm:preparation} is an equality of states for each $M$, and the ancillary registers end in the same state $\ket{\Gamma_d}_{\mathrm{bond}}\ket{0}_M$ for every $M$. By linearity, if the circuit is initialised with the $M$ register in a superposition $\sum_M\beta_M\ket{M}$, it prepares
\begin{equation}\label{eq:m-superposition}
    \ket{\Gamma_d}_{\mathrm{bond}}\otimes\ket{0}_M\otimes\sum_{M=-S}^{S} \beta_M\,\ket{\Psi_{SM}},
\end{equation}
which is an arbitrary state of the multiplet. Second, while we suggest interleaving the site isometries and inverse Clebsch-Gordan transformations, the lower row of Equation~\ref{eq:diag-two-layer} can also be applied on its own, in serial order after the whole step-vector MPS has been prepared by Corollary~\ref{corollary:loading}.
The spin would then be kept in a separate register that starts at $S$ and is updated at each site from the step, and the inverse Clebsch-Gordan transforms of all the sites compose to the inverse quantum Paldus transform of Ref.~\cite{burkat2025quantum}, without its particle-number register and with its qubit order reversed because of our right-to-left spin fusion convention. This two-stage circuit prepares the same state for every $M$. The circuit of Theorem~\ref{thm:preparation} interleaves the two stages, so that $S_i$ can be read from the bond register and no spin register is needed; Section~\ref{sec:cost} compares the costs of the two options, which are not significantly different.

\subsection{Cost}\label{sec:cost}

By Corollary~\ref{corollary:loading}, the site isometries inherit the cost for block-sparse preparation with the protocol of Ref.~\cite{rupprecht2026faster}, except that it is instead for blocks of the site matrix of size $h^{(i)}_{\Gamma}\times n^{(i-1)}_{\Gamma}$, i.e. sizes that count multiplets rather than states of the right block. Preparing the same state from its non-spin-adapted form at a single projection $M$ involves blocks whose sizes count states instead. For a singlet, bond $i$ of the non-spin-adapted MPS carries $\chi^{\ast}_i=\sum_{\Gamma}n^{(i)}_{\Gamma}(2S_\Gamma+1)$ states, whereas the spin-adapted MPS carries $\chi_i=\sum_{\Gamma}n^{(i)}_{\Gamma}$ multiplets (Corollary~\ref{corollary:loading}(ii)). This difference is the saving of our method, and we will shortly quantify it for concrete examples in Section~\ref{sec:numerics}.

The inverse Clebsch-Gordan transforms add a negligible cost that does not depend on the entries of the tensors (Theorem~\ref{thm:preparation}(iii)). The rotation at site $i$ is multiplexed over at most $\tfrac12(k+1)(k+2)$ pairs $(S_i,M_{i-1})$, with $k=d+1-i$~\cite{burkat2025quantum}, and the $M$ register has $n_M=\lceil\log_2(2d+1)\rceil$ qubits, so the rotation and the update of $M$ depend solely on the number of orbitals. Only the read-out of $S_i$ depends on the bond dimension. It is free if the bond register stores the sector explicitly, and is otherwise a table lookup over the $\chi_i$ values of the bond index. The two-stage circuit of Section~\ref{sec:preparation} replaces the read-out by a spin register of $n_S=\lceil\log_2(d+1)\rceil$ qubits, updated at each site with $6(n_S+1)$ Toffoli gates~\cite{burkat2025quantum}. In the examples of Section~\ref{sec:numerics} the transforms, including read-out, account for less than $0.2\%$ of the Toffoli gates, and the difference between the two circuits is smaller still, so the choice between them is largely immaterial.

\section{Numerical experiments}\label{sec:numerics}
The cost of the state-preparation circuit depends on the state through its spin-adapted bond dimension $\chi$, counted in multiplets and resolved by symmetry sector (Corollary~\ref{corollary:loading}), which is known only once a DMRG calculation has been run. We therefore estimate the cost of the circuit for initial states of two popular benchmark systems, and compare it with the dense \cite{berry2025rapid} and block-sparse \cite{rupprecht2026faster} preparation routines on the same states.

Our examples are the singlet ground states of complete-active-space (CAS) models of the iron-sulphur clusters, that is the CAS(30e,\,20o) model of [Fe$_2$S$_2$]$^{2-}$, and the CAS(54e,\,36o) model of [Fe$_4$S$_4$]$^{2-}$.
The strong correlation of iron-sulphur clusters has made them, and the FeMo cofactor of nitrogenase above all, a standard target for quantum chemistry on quantum computers~\cite{sharma2014low,li2017spin,reiher2017elucidating,berry2019qubitization,lee2021even,lee2023evaluating,low2025fast,robledo2025chemistry}, and these two clusters a common benchmark for initial-state preparation~\cite{lee2023evaluating,fomichev2024initial,ollitrault2024enhancing,marti2025spin,berry2025rapid,rupprecht2026faster}.

\subsection{State generation and cost model}\label{sec:numerics-setup}
To facilitate direct comparison, we take the integrals of both models from the data record of Ref.~\cite{rupprecht2026faster}, itself related to those of Refs.~\cite{sharma2014low,li2017spin,lee2023evaluating,ollitrault2024enhancing}.
For each model we ran a spin-adapted DMRG calculation with \texttt{block2}~\cite{zhai2023block2} at a bond dimension of $\chi=1000$ multiplets, following the script published alongside Ref.~\cite{rupprecht2026faster}; the full details of the DMRG calculation are given in Appendix~\ref{app:dmrg}.
We count Toffoli gates and qubits using the code published alongside Ref.~\cite{rupprecht2026faster}, itself built upon \texttt{Qualtran}~\cite{harrigan2024expressing}. The compared protocols are costed as published, hence the dense baseline does not include the real-tensor saving introduced in Ref.~\cite{rupprecht2026faster}, but the $\U1\times \U1$-adapted baseline does. We assume 15-bit rotation angles, a generous allocation relative to the DMRG truncation error. The code estimates the cost of the site isometries of all three methods, but not of the Clebsch-Gordan transform (used only in our work), so we built \texttt{Qualtran} bloqs for the inverse transform of Theorem~\ref{thm:preparation} from standard lookup, rotation and addition bloqs (i.e. at each site a lookup of $S_i$ from the bond index, a lookup of the rotation angle addressed by $S_i$ and $M_{i-1}$, the rotation, and two controlled additions to the $M$ register). The angle lookup covers only the spins present on the bond.
Each table lookup is a QROAM and therefore can trade Toffoli gates for ancilla qubits~\cite{berry2019qubitization,low2024trading}. In all our counts, for the baselines as well as for our circuit, every lookup is set to minimise its Toffoli count, with no limit on the ancilla qubits. Ref.~\cite{rupprecht2026faster} instead limited the ancilla qubits of some lookups by hand, leading to some minor differences in ancilla counts reported between our works.

\subsection{Comparisons}\label{sec:numerics-comparison}
Existing leading methods for fault-tolerant MPS preparation load the state in its non-spin-adapted form, using either a dense representation~\cite{berry2025rapid} (the dense preparation) or a block-sparse representation~\cite{rupprecht2026faster} (the $\U1\times\U1$ preparation). For a fair comparison, we compare our protocol to both of these in two ways: for the same state, and at equal fidelity after truncation.

For the same-state comparisons, we convert each spin-adapted MPS to its non-spin-adapted form (Equation~\eqref{eq:site-factorisation}), taken at $M=0$, and count both preparations of it with the cost model above.
The non-spin-adapted form can sometimes be truncated to a smaller bond dimension at little loss of fidelity~\cite{rupprecht2026faster}, and a truncated state is accordingly cheaper to prepare, which is why we also compare at equal fidelity. We truncate both forms of each state to a range of bond dimensions, counted in multiplets ($\chi$) for the spin-adapted form and in states ($\chi^{\ast}$) for the non-spin-adapted form, keeping at every bond the largest Schmidt values across all sectors. 
For the spin-adapted form, the Schmidt values are those of whole multiplets: a multiplet of spin $S_\Gamma$ with Schmidt value $\lambda$ carries the weight $\lambda^2$ in the state, and appears in the non-spin-adapted form as $2S_\Gamma+1$ states with Schmidt value $\lambda/\sqrt{2S_\Gamma+1}$ each (see proof of Corollary~\ref{corollary:loading}(ii)).
We then compute, for each truncated state $|\tilde{\Psi}\rangle$, the cost of preparing it and its \emph{truncation infidelity} $1-|\langle\Psi|\tilde{\Psi}\rangle|^2$ with respect to the untruncated state $\ket{\Psi}$.
We remark that different results might be obtained in the abelian case if using DMRG adapted to $\U1\times\U1$ symmetry, since that permits a state that is not an eigenstate of total spin; our comparison of a spin-adapted state to its non-spin-adapted expansion is in line with that of Ref.~\cite{rupprecht2026faster}.

\subsection{Results and discussion}\label{sec:numerics-results}
\begin{table*}[t]
\caption{\label{tab:costs}\textbf{Toffoli costs of preparing each state with the three methods} (and ancilla counts in parentheses): `Dense' is the preparation of the non-spin-adapted form without symmetry as published in Ref.~\cite{berry2025rapid}; `$\U1\times\U1$' is the preparation of the non-spin-adapted form exploiting abelian block-sparsity~\cite{rupprecht2026faster}; and `This work' is the preparation of the spin-adapted form using our protocol, per Theorem~\ref{thm:preparation}.}
\begin{ruledtabular}
\begin{tabular}{lcccccccc}
System & $\chi$ & $\chi^{\ast}$ & Dense~\cite{berry2025rapid} & $\U1\times\U1$~\cite{rupprecht2026faster} & \multicolumn{2}{c}{This work} & \multicolumn{2}{c}{Ratio to this work} \\
\cline{6-7}\cline{8-9}
 & & & & & Total & Clebsch-Gordan & Dense & $\U1\times\U1$ \\
\hline
{}[Fe$_2$S$_2$]$^{2-}$ & 1000 & 3299 & $8.32\times10^{7}$ (516) & $5.46\times10^{6}$ (514) & $1.63\times10^{6}$ (267) & $2.56\times10^{3}$ & 51 & 3.4 \\
{}[Fe$_4$S$_4$]$^{2-}$ & 1000 & 7251 & $5.52\times10^{8}$ (997) & $2.00\times10^{7}$ (996) & $4.09\times10^{6}$ (268) & $6.36\times10^{3}$ & 135 & 4.9 \\
\end{tabular}
\end{ruledtabular}
\end{table*}

\begin{figure}[t]
\centering
\includegraphics[width=\columnwidth]{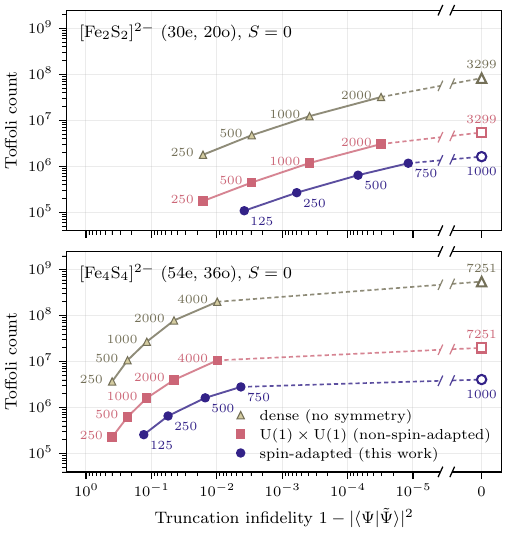}
\caption{\label{fig:cost-fidelity}\textbf{Comparing Toffoli costs at varying truncation fidelities.} The spin-adapted preparation of this work (circles, blue) is compared with the $\U1\times\U1$ preparation~\cite{rupprecht2026faster} (squares, red) and the dense preparation (triangles, gold) of the non-spin-adapted form~\cite{berry2025rapid}. Each point shows the state truncated to the bond dimension written next to it, in multiplets ($\chi$) for the spin-adapted case and in states ($\chi^{\ast}$) for the non-spin-adapted cases. The open symbols denote the untruncated states (i.e. zero truncation infidelity).}
\end{figure}

We now apply the cost model of Section~\ref{sec:numerics-setup} to the two Fe-S clusters and make the two comparisons of Section~\ref{sec:numerics-comparison} in turn.
Table~\ref{tab:costs} compares the three preparations on the same untruncated state, and Figure~\ref{fig:cost-fidelity} compares them when each form of the state is truncated to a given fidelity.
We then show that the inverse Clebsch-Gordan transforms contribute very little to these costs, and trace the saving to the spins of the multiplets on the bonds.

For the untruncated states, the spin-adapted preparation needs up to $4.9\times$ fewer Toffoli gates than the $\U1\times\U1$ preparation~\cite{rupprecht2026faster}, and up to $135\times$ fewer than the dense preparation~\cite{berry2025rapid}. In detail (Table~\ref{tab:costs}), the saving over the $\U1\times\U1$ preparation is a factor of 3.4 for [Fe$_2$S$_2$]$^{2-}$ and of 4.9 for [Fe$_4$S$_4$]$^{2-}$, and the saving over the dense preparation is a factor of 51 and of 135 respectively. The spin-adapted preparation also needs fewer ancilla qubits: about half as many as either of the other two preparations for [Fe$_2$S$_2$]$^{2-}$, and about a quarter as many for [Fe$_4$S$_4$]$^{2-}$. These savings reflect the bond dimensions in Table~\ref{tab:costs}: both preparations of the non-spin-adapted form load tensors whose bonds hold up to $\chi^{\ast}=3299$ and $\chi^{\ast}=7251$ states, whereas the bonds of the spin-adapted MPS hold at most $\chi=1000$ multiplets.

However, the comparison on untruncated states would overstate our advantage if the non-spin-adapted form could be truncated much further than the spin-adapted form at the same loss of fidelity (Section~\ref{sec:numerics-comparison}), a possibility raised by Ref.~\cite{rupprecht2026faster}. Figure~\ref{fig:cost-fidelity} shows that against the $\U1\times\U1$ preparation this is not the case for either cluster: the saving at equal fidelity is at least as large as for the untruncated states. At equal truncation infidelity, the spin-adapted preparation needs $3.5$--$3.6\times$ fewer Toffoli gates than the $\U1\times\U1$ preparation for [Fe$_2$S$_2$]$^{2-}$, and $5.0$--$5.4\times$ fewer for [Fe$_4$S$_4$]$^{2-}$. Against the dense preparation the corresponding improvement factors are 37 to 38 for [Fe$_2$S$_2$]$^{2-}$ and 91 to 95 for [Fe$_4$S$_4$]$^{2-}$. These ratios are taken at the spin-adapted states of $\chi=125$, $250$ and $500$, with the non-spin-adapted series interpolated linearly in log-space between its points, and therefore are approximate.
The same data can be read at equal cost: at the same Toffoli count, the spin-adapted preparation reaches a higher fidelity, which would raise the success probability of QPE and reduce the number of repetitions it needs.

The inverse Clebsch-Gordan transforms are the one part of our circuit that the other two preparations do not need, and their cost is negligible in comparison to loading the reduced-tensor amplitudes themselves.
For the untruncated states, they account for less than $0.2\%$ of our Toffoli count (Table~\ref{tab:costs}), and they need at most 74 and 91 ancilla qubits, well below the peak costs of the site isometries.
Their cost is also almost independent of the bond dimension: between $\chi=125$ and $\chi=1000$ it grows only by $27\%$ and $40\%$, whereas that of the site isometries grows about fifteenfold, so their share appears to fall as the bond dimension grows. The cost of our circuit is therefore overwhelmingly dominated by that of its site isometries, and is almost as close to parameter-optimal as the current state-of-the-art in fault-tolerant unitary synthesis permits.

The size of the Toffoli saving depends on the spins of the multiplets on the bonds, which is why the saving is larger for [Fe$_4$S$_4$]$^{2-}$ than for [Fe$_2$S$_2$]$^{2-}$. For a singlet state, a multiplet of spin $S_\Gamma$ on a bond of the spin-adapted MPS corresponds to $2S_\Gamma+1$ states on the same bond of the non-spin-adapted form (Section~\ref{sec:cost}), which share its weight equally (Section~\ref{sec:numerics-comparison}). A truncation of the non-spin-adapted form therefore needs $2S_\Gamma+1$ states for every multiplet it keeps, and it drops multiplets of high spin first. The more of the Schmidt weight is carried by multiplets of high spin, the larger the saving from spin adaptation.

In both clusters these spins are large, although the total spin is zero: at the middle bond, $83\%$ of the weight $\lambda^2$ is in multiplets of spin $2$ and $5/2$ for [Fe$_2$S$_2$]$^{2-}$, and $80\%$ is in multiplets of spin $4$ and $9/2$ for [Fe$_4$S$_4$]$^{2-}$. Both clusters are thus favourable cases for spin adaptation, since both have large spins on their bonds. For a multiplet target of total spin $S>0$ the comparison would additionally depend on the projection $M$ at which the non-spin-adapted form is taken, and on whether the bonds are described by the left or the right block. The circuit remains to be applied to a wider range of molecules, active spaces, spin states and bond dimensions.

\section{Conclusions}\label{sec:conclusion}
In this work, we have shown that the obstacle to preparing a spin-adapted MPS at the cost of its classical description was predominantly one of labelling.
Total spin does not simply add up from one orbital to the next in the occupation basis, but once each site is labelled by its step of the unitary group approach, the reduced tensors form an abelian-symmetric MPS that the block-sparse protocol of Ref.~\cite{rupprecht2026faster} can load without modification.
The non-abelian symmetry then requires only a fixed change of basis, the inverse of the quantum Paldus transform~\cite{burkat2025quantum}, at negligible cost.
The compression that spin adaptation brings to the classical calculation is thereby inherited by the quantum circuit. The saving is set by the spins of the multiplets on the bonds, so it can be large even for a target of zero total spin: both of our examples are singlets.

The remaining gap between the cost of our circuit and the number of parameters that the classical calculation stores arises from synthesis, not from symmetry. As Rupprecht and W\"{o}lk observe~\cite{rupprecht2026faster}, completing each rectangular block to a unitary loads rotation angles that a parameter count shows are not needed; synthesising the blocks directly as isometries could remove this overhead, for their preparation as much as for ours.
Abelian point-group symmetries of the Hamiltonian can easily be incorporated in our protocol by adding one more label to the blocks (Section~\ref{sec:loading}).

Finally, we note that very little in our construction is specific to $\SU2$. It relies on the fact that each total spin occurs at most once when two spins are coupled, so that each block of a reduced tensor belongs to exactly one step. We therefore expect the same two stages to carry over to MPS adapted to other non-abelian symmetries: a block-sparse loading over steps, followed by a state-independent cascade of Clebsch-Gordan transforms as in the quantum Schur transform \cite{bacon2005quantum}.
The steps, and the transforms that undo them, are then dictated by the representation theory of the group in question.
This will of course not be automatic, since for larger groups such as $\SU3$ the same symmetry label can occur more than once in a coupling, and the step would have to record which.

\section*{Related work}
During final preparation of this manuscript, we became aware of an independent pre-print released several days ago~\cite{hennefarth2026quantum} which introduces a state-preparation protocol based on similar principles (i.e. preparing a spin-adapted MPS directly from its reduced tensors). We encourage the interested reader to consult both works for a complete picture.

\section*{Acknowledgements}
We acknowledge helpful discussions with Tom Bromley and Annina Lieberherr. This research was supported by The University of Melbourne’s Research Computing Services and the Petascale Campus Initiative. This work was part of the QUICHE project (QUantum Integrated CHEmistry, reference: 10150101) funded by Innovate UK.

\subsection*{AI usage statement}
Generative AI tools (Claude Opus 4.8, 5 \& 5.5, Claude Fable 5, 5.1 \& 5.2) were used under close supervision throughout the course of the project to assist with the following: recommendations and feedback on drafts, formatting of equations, mathematical statements and diagrams, exploratory work on theoretical derivations, and coding tasks for numerical experiments. All outputs from these tools that affect the final manuscript were critically reviewed line-by-line by the authors, who take full responsibility for its content and accuracy.

\clearpage
\newpage
\widetext

\appendix
\section*{Appendices for ``Spin-adapted state preparation for quantum chemistry''}

Here we present additional details for the main results in our manuscript, structured as follows.
In Appendix~\ref{app:conventions}, we present full detail on taking the input tensors from a spin-adapted DMRG code, with regard to both the sign and normalisation conventions that the tensors must satisfy (Appendix~\ref{app:signs}) and the DMRG calculations behind the numerical results of Section~\ref{sec:numerics} (Appendix~\ref{app:dmrg}).
In Appendix~\ref{app:floating}, we outline the floating-bond construction for MPS preparation, in which the bond register is placed on the qubits not yet prepared, sparing the use of dedicated bond ancillae.

\section{Input from spin-adapted DMRG codes}\label{app:conventions}

\subsection{Sign conventions and normalisation}\label{app:signs}

Three conventions fix the signs of individual amplitudes in Equation~\eqref{eq:su2-states}:
(i) the coupling order in the Clebsch-Gordan coefficient: the right block comes first and the orbital second, with the Condon-Shortley phase of Definition~\ref{def:cg}; (ii) the operator order of Equation~\eqref{eq:block-product}: the creation operators of orbital $i$ stand to the left of those of the right block, so that every basis vector is $A^\dagger_1 A^\dagger_2\cdots A^\dagger_d\ket{\bm{0}}$ with the orbital index increasing to the right, in the interleaved Jordan-Wigner order of Section~\ref{sec:problem-setup}; (iii) the doubly occupied state $a^\dagger_{\uparrow}a^\dagger_{\downarrow}\ket{\emptyset}$ of Equation~\eqref{eq:local-multiplets}. These are the conventions of this work. Lemma~\ref{lemma:mapping} relies on them: with these choices, the amplitudes of the step-vector MPS are the coefficients of the state in the CSF basis of Section~\ref{sec:uga}.

These conventions matter when the tensors are taken from a DMRG library, which may make different choices; conventions differ significantly for spin-adapted codes especially. We account for the conventions of the library \texttt{block2}~\cite{zhai2023block2}, used in Section~\ref{sec:numerics}. A \texttt{block2} calculation whose last sweep runs right-to-left, leaving the orthogonality centre at site~$1$, stores its tensors as our paper assumes: right-canonical, with the quantum numbers of the right block on the bonds. Read in this way, \texttt{block2} agrees with (ii) and (iii) but couples the orbital first and the right block second, the reverse of (i).

Reversing any one of the three conventions, or all three, multiplies every amplitude by $(-1)$ to the number of its negatively spin-coupled steps $\delta_i=2$, up to a global sign. Reversing two of them is a gauge transformation on the bonds times a global sign, and leaves the state invariant. The tensors sourced from \texttt{block2} are therefore brought to the conventions of this work by multiplying every block of a negatively spin-coupled step by $-1$, and need no other change (so long as DMRG was concluded on a right-to-left sweep).
When preparing multiplet states ($S>0$), there is a second reason to end on a right-to-left sweep: the number of multiplets on a bond differs between its description by the left block and by the right block (Section~\ref{sec:su2-mps}), and the cost of our circuit is set by the latter, so the bond dimension $\chi$ must be imposed in the right-block description.

For any other library, these conventions must be checked carefully before its tensors are loaded, and so does the normalisation: a stored tensor is the reduced tensor of Equation~\eqref{eq:site-factorisation} only if the full element is recovered from it by exactly one Clebsch-Gordan coefficient, so a sector-dependent factor such as $\sqrt{2S_i+1}$ that a code absorbs into its tensors must be multiplied out before loading.

\subsection{The DMRG calculations of this work}\label{app:dmrg}

The two states studied in Section~\ref{sec:numerics} come from spin-adapted DMRG calculations with \texttt{block2}~\cite{zhai2023block2} that follow the script published alongside Ref.~\cite{rupprecht2026faster}. They use the two-site  DMRG algorithm with point-group symmetry disabled, and start from a random MPS. The orbitals are placed in Fiedler order, which places orbitals with large exchange integrals close together on the chain~\cite{zhai2023block2}. Table~\ref{tab:dmrg} gives the sweep schedule, which ends at $\chi=1000$ multiplets. The last sweep must run from right to left, so that the stored tensors are of the form that the circuit takes (Appendix~\ref{app:signs}); where the schedule ends in the other direction, we add one noiseless sweep to restore the conventions of this paper.

Table~\ref{tab:dmrg} also gives the energy and the discarded weight of each final state. The [Fe$_4$S$_4$]$^{2-}$ calculation reaches the end of the schedule without meeting its convergence criterion.
Convergence is not the aim for QPE guide state preparation, unlike the use of DMRG for direct computation of ground states. QPE is needed precisely for the systems on which DMRG at an affordable bond dimension does not converge, since those on which it does are already solved classically, and there the DMRG state serves through its overlap with the ground state. The comparison of Section~\ref{sec:numerics} is in any case unaffected, and it is the multiplets, their sectors and multiplicities that determine the state-preparation cost, not the minutiae of their amplitudes themselves.
\begin{table}[H]
\caption{\label{tab:dmrg}\textbf{The DMRG calculations.} (a) The sweep schedule, from the script published with Ref.~\cite{rupprecht2026faster}. $\chi$ is the largest number of multiplets kept on a bond, and the sweeps stop early once the energy changes by less than $10^{-7}$\,Ha from one sweep to the next. (b) The states in which the calculations end. The discarded weight is the largest of the last sweep.}
\renewcommand{\arraystretch}{1.2}
\begin{minipage}[t]{0.43\textwidth}
\centering
(a)\\[0.6ex]
\begin{tabular}[t]{c@{\hspace{1.8em}}c@{\hspace{1.8em}}c@{\hspace{1.8em}}c}
\hline\hline
Sweeps & $\chi$ & Noise & Davidson threshold \\
\hline
1--4 & 200 & $10^{-4}$ & $10^{-4}$ \\
5--8 & 400 & $10^{-4}$ & $10^{-4}$ \\
9--12 & 600 & $10^{-4}$ & $10^{-4}$ \\
13--16 & 800 & $10^{-4}$ & $10^{-4}$ \\
17--20 & 1000 & $10^{-4}$ & $10^{-4}$ \\
21--24 & 1000 & $10^{-5}$ & $10^{-5}$ \\
25--28 & 1000 & $10^{-6}$ & $10^{-6}$ \\
29--32 & 1000 & $10^{-7}$ & $10^{-7}$ \\
33--38 & 1000 & 0 & $10^{-8}$ \\
\hline\hline
\end{tabular}
\end{minipage}\hfill
\begin{minipage}[t]{0.55\textwidth}
\centering
(b)\\[0.6ex]
\begin{tabular}[t]{l@{\hspace{1.2em}}c@{\hspace{1.2em}}c}
\hline\hline
 & [Fe$_2$S$_2$]$^{2-}$ & [Fe$_4$S$_4$]$^{2-}$ \\
\hline
Energy (Ha) & $-5092.871026$ & $-8432.791493$ \\
Discarded weight & $1.3\times10^{-6}$ & $1.4\times10^{-4}$ \\
Energy change, last sweep (Ha) & $-6\times10^{-9}$ & $-5\times10^{-7}$ \\
\hline\hline
\end{tabular}
\end{minipage}
\end{table}

\section{Floating-bond MPS preparation}\label{app:floating}

The sequential MPS preparation scheme in Refs.~\cite{berry2025rapid,rupprecht2026faster} utilises a dedicated bond register of $w=\lceil\log_2\chi\rceil$ qubits alongside the $2d$ system qubits. Here we note that the bond register may instead `float' along the $2d$ uninitialised system qubits, sparing the use of $w$ dedicated ancillae. The saving of $w$ qubits is small, but in a space-constrained compilation it is not idle: QROM-based loading converts spare logical qubits into a lower Toffoli count~\cite{berry2019qubitization,low2024trading}.

We seek to prepare a matrix product state $\ket{\Psi}$, Equation~\eqref{eq:mps}, optimised via DMRG. The MPS has $d$ sites of dimension $D$ ($D=4$ for a spatial orbital in the standard occupation basis $\{\ket{\emptyset},\ket{\uparrow},\ket{\downarrow},\ket{\uparrow\downarrow}\}$), where $M_i^{\sigma}$ has shape $\chi_{i-1}\times\chi_i$ with $\chi_0=\chi_d=1$, and the tensors are chosen to be in right-canonical gauge, Equation~\eqref{eq:right-canonical} (such that the tensors are isometries implementable with unitary operations). In this appendix we number the physical indices $\sigma\in\{0,\dots,D-1\}$ and the bond indices $\alpha_i\in\{0,\dots,\chi_i-1\}$ from zero. Writing $w_i\equiv\lceil\log_2\chi_i\rceil$ for the number of qubits needed to hold the bond index at cut $i$ (therefore $w_0=w_d=0$), we note that the floating-bond register will always fit within the uninitialised system qubits:

\begin{proposition}[Fit bound]\label{prop:fit}
Let $p\equiv\lceil\log_2 D\rceil$ be the number of qubits per site and $w_i\equiv\lceil\log_2\chi_i\rceil$. Then (i) $w_{i-1}\le w_i+p$ and (ii) $w_i\le p(d-i)$; i.e.\ the bond index at cut $i$ always fits within the uninitialised system qubits of sites $i+1,\dots,d$.
\end{proposition}
\begin{proof}
The $\chi_{i-1}\times D\chi_i$ matrix $[M_i^{0}\cdots M_i^{D-1}]$ has orthonormal rows by right-canonicality, Equation~\eqref{eq:right-canonical}, so $\chi_{i-1}\le D\chi_i\le 2^p\chi_i$, implying (i); iterating from the right edge gives $\chi_i\le D^{d-i}\chi_d$, and $w_i\le\lceil(d-i)\log_2 D\rceil\le p(d-i)$ yields (ii).
\end{proof}

From this point, preparing the MPS with the floating-bond construction is a simple case of applying the $M_i^{\sigma}$ isometries on the correct bonds with the correct shaping. We denote the binary encoding of $\alpha\in\{0,\dots,\chi_i-1\}$ (zero-padded to $w_i$ bits, basis states of any block read as integers least-significant-bit first) as $\mathrm{enc}(\alpha)\in\{0,1\}^{w_i}$, and label the $pd$ system qubits $q_1,\dots,q_{pd}$ in Jordan-Wigner order, site $i$ mapping to $Q_i\equiv(q_{p(i-1)+1},\dots,q_{pi})$. Then, the state preparation acts on a rolling `window' $W_i\equiv(q_{p(i-1)+1},\dots,q_{pi+w_i})$, the left side of which shifts $p$ qubits to the right per isometry and the right side of which moves (or stays stationary) according to $\chi_i$, dynamically adapting the window to the amount of entanglement information in that bond. More precisely, we must construct unitaries $U_1,\dots,U_d$ such that
\begin{equation}\label{eq:floating-invariant}
    U_i\cdots U_1\ket{0}^{\otimes pd} = \sum_{\sigma_1\dots\sigma_i}\sum_{\alpha_i=0}^{\chi_i-1}\big(M_1^{\sigma_1}\cdots M_i^{\sigma_i}\big)_{0,\alpha_i}\ket{\sigma_1}_{Q_1}\cdots\ket{\sigma_i}_{Q_i}\otimes\ket{\mathrm{enc}(\alpha_i)}_{q_{pi+1}\dots q_{pi+w_i}}\otimes\ket{0}^{\otimes(pd-pi-w_i)}.
\end{equation}
Applying iteratively to $\ket{0}^{\otimes pd}$, we see that this is achieved by unitaries that map
\begin{equation}\label{eq:floating-site}
    U_i:\quad\ket{\mathrm{enc}(\alpha_{i-1})}\ket{0\cdots0}\longmapsto\sum_{\sigma_i}\sum_{\alpha_i=0}^{\chi_i-1}\big(M_i^{\sigma_i}\big)_{\alpha_{i-1},\alpha_i}\ket{\sigma_i}_{Q_i}\ket{\mathrm{enc}(\alpha_i)}_{q_{pi+1}\dots q_{pi+w_i}}.
\end{equation}
The incoming encoding $\ket{\mathrm{enc}(\alpha_{i-1})}$ occupies the leading $w_{i-1}\le w_i+p$ qubits of $W_i$ (Proposition~\ref{prop:fit}(i)) with the remaining window qubits in $\ket{0}$, while the outgoing encoding occupies the leading qubits of $W_{i+1}$; the maps therefore compose as claimed, the floating bond register never overflows outside the $pd$ system qubits (Proposition~\ref{prop:fit}(ii)), and from Equations~\eqref{eq:floating-invariant} and~\eqref{eq:mps} we conclude $U_d\cdots U_1\ket{0}^{\otimes pd}=\ket{\Psi}$.

Finally, we note that per Equation~\eqref{eq:floating-site}, enacting $U_i$ requires only the synthesis of the first $\chi_{i-1}$ columns of a $2^{p+w_i}$-dimensional unitary, whose column $\alpha$ is precisely the reshaped tensor
\begin{equation}\label{eq:floating-cols}
    \tilde V_i[\sigma+2^p\alpha',\alpha] := (M_i^{\sigma})_{\alpha\alpha'},
\end{equation}
that is, the isometry $[M_i^{0}\cdots M_i^{D-1}]^T$ of Ref.~\cite{berry2025rapid} with rows re-indexed classically so that the physical label occupies the $p$ lowest bits. Since this re-indexing only relabels rows and columns, a block-sparse $M_i$ keeps its blocks and their sizes, and the procedure therefore maps exactly to that of Section~III in Ref.~\cite{berry2025rapid}, including the block-sparse variant of Ref.~\cite{rupprecht2026faster}, with the blocks of Corollary~\ref{corollary:loading} for the step-vector MPS. We note that this does not by itself change Toffoli counts compared to implementations employing a dedicated bond register: such a register can apply the same $(p+w_i)$-qubit unitaries $U_i$ once a network of SWAP gates, which costs no Toffoli gates, has rotated the incoming index into place. The savings for tensors near the edge that do not fully saturate the bond dimension $\chi$, where the QROM costs of outputting amplitudes for $\chi_i\le\chi$ are lowered (e.g.\ in particular, the final isometry reduces from a lookup over $D\chi$ entries to a $p$-qubit unitary), are therefore available to both layouts; relative to a careful implementation of the standard construction this is purely a saving of $\lceil \log_2 \chi \rceil$ ancilla qubits.

By Corollary~\ref{corollary:loading}, the site isometries of Theorem~\ref{thm:preparation} are those of the block-sparse preparation of a right-canonical step-vector MPS, so the floating-bond construction applies to them unchanged ($D=4$, $p=2$ and the steps in the code of Section~\ref{sec:cg-step}). The inverse Clebsch-Gordan transforms $\hat K_{S_i}$ act only on $Q_i$ and the $M$ register, with $S_i$ read from the bond index on the qubits that follow $Q_i$, so the circuit of Theorem~\ref{thm:preparation} prepares $\ket{\Psi_{SM}}\otimes\ket{0}_M$ on $2d+n_M$ qubits besides workspace, a modest saving of $\lceil\log_2\max_i\chi_i\rceil$ fewer than with a bond register of its own. To make the read-out of $S_i$ free, as described in Section~\ref{sec:cost}, the spin should be kept in a dedicated ancilla register rather than in the floating bond register: the spin register of Section~\ref{sec:cost}, $n_S=\lceil\log_2(d+1)\rceil$ qubits holding a copy of $S_i$, updated from the step at every site. Storing the sector as an extra field of the floating bond index instead can make it wider than $w_i$, so that it no longer fits within the uninitialised qubits. Without a spin register, $S_i$ is read out of $\mathrm{enc}(\alpha)$ by the table lookup of Section~\ref{sec:cost}.

\end{document}